\documentclass[10pt, twoside, twocolumn]{IEEEtran}
 \makeatletter

\newcommand{\Rmnum}[1]{\expandafter\@slowromancap\romannumeral #1@}
\makeatother
\usepackage{graphicx,cite,epsfig,amssymb,amsmath,multirow}
\usepackage{amsfonts}
\usepackage{mathrsfs}
\usepackage{algorithm}
\usepackage{algorithmic}
\usepackage{float}
\usepackage{bm}
\usepackage{booktabs}
\usepackage{epstopdf}
\usepackage{ulem}
\usepackage{slashbox}
\usepackage{subfigure}
\usepackage[usenames,dvipsnames]{color}
\usepackage{color}

\newtheorem{lemma}{{Lemma}}
\newtheorem{corollary}{{Corollary}}

\makeatother
\begin{document}
\bibliographystyle{ieeetr}

\title{Spatial Lobes Division Based Low Complexity Hybrid Precoding and Diversity Combining for mmWave IoT Systems }

\author{Yun~Chen,~Da~Chen,~Yuan~Tian,~and~Tao~Jiang, \emph{Senior Member, IEEE}
\thanks{Manuscript received September 7, 2018; revised November 3, 2018; accepted November 6, 2018; Date of publication ...; date of current version ... This work was supported in part by the National Science Foundation of China with Grant numbers 61771216, 61631015 and 61729101, Fundamental Research Funds for the Central Universities with Grant number 2015ZDTD012, China Scholarship Council (CSC), and the National Science Foundation of China with Grant 61601191. (\emph{Corresponding author: Tao Jiang.})

Y.~Chen,~D.~Chen,~Y.~Tian,~and~T.~Jiang are with Wuhan National Laboratory for Optoelectronics, School of Electronic Information and Communications, Huazhong University of Science and Technology, Wuhan 430074, China (e-mail: chen\_yun@hust.edu.cn; chenda@hust.edu.cn; yuan\_tian@hust.edu.cn; tao.jiang@ieee.org).

Copyright (c) 2012 IEEE. Personal use of this material is permitted. However, permission to use this material for any other purposes must be obtained from the IEEE by sending a request to pubs-permissions@ieee.org.
}}

%\IEEEpubid{Copyright (c) 2012 IEEE. Personal use of this material is permitted.
%However, permission to use this material for any other purposes must be obtained from the IEEE by sending a request to pubs-permissions@ieee.org.}

\markboth{IEEE Internet of Things Journal, ~Vol.~, No.~, 2018}%
{Chen \MakeLowercase{\textit{et al.}}: SLD based Low Complexity HYP and Diversity Combining for mmWave IoT Systems }

\maketitle

\begin{abstract}
This paper focuses on the design of low complexity hybrid analog/digital precoding and diversity combining in millimeter wave (mmWave) Internet of things (IoT) systems. Firstly, by exploiting the sparseness property of the mmWave in the angular domain, we propose a spatial lobes division (SLD) to group the total paths of the mmWave channel into several spatial lobes, where the paths in each spatial lobe form a low-rank sub-channel. Secondly, based on the SLD operation, we propose a low complexity hybrid precoding scheme, named HYP-SLD. Specifically, for each low-rank sub-channel, we formulate the hybrid precoding design as a sparse reconstruction problem and separately maximizes the spectral efficiency. Finally, we further propose a maximum ratio combining based diversity combining scheme, named HYP-SLD-MRC, to improve the bit error rate (BER) performance of mmWave IoT systems. Simulation results demonstrate that, the proposed HYP-SLD scheme significantly reduces the complexity of the classic orthogonal matching pursuit (OMP) scheme. Moreover, the proposed HYP-SLD-MRC scheme achieves great improvement in BER performance compared with the fully digital precoding scheme.
\end{abstract}


\section{Introduction}
\IEEEPARstart{T}{he} wireless data traffic of Internet of things (IoT) is expected to grow exponentially in the next few years, since the number of access devices in IoT will face explosive growth \cite{Boxi2017}. Millimeter wave (mmWave) technology, which is one of the most important technologies for IoT, can greatly alleviate the above traffic pressure due to the abundant spectrum resources and large bandwidth \cite{Kong2017}. Moreover, the small wavelength of mmWave signals enables the deployment of large antenna arrays into small space of IoT devices \cite{4Heath2016}. By now, the 60 GHz mmWave communication protocols have been introduced in IEEE 802.11ad and 802.11ay \cite{Ghasempour2017, Silva2018}. However, owing to the high carrier frequency, mmWave signals experience severe attenuation, which makes mmWave face many challenges when applied to the IoT \cite{Hemadeh2018}. As another promising technology for IoT and 5G, massive multiple-input multiple-output (MIMO) could generate high precoding gains to compensate for the high path loss of mmWave through the precoding technology \cite{Ticao2016, PLiu2017, SQiu2018, XBWang2017}. Therefore, it is of great significance to study precoding schemes in mmWave IoT systems \cite{6Rappaport2014}.

There are three main candidate precoding schemes, i.e., fully digital precoding, analog precoding and hybrid precoding. The fully digital precoding is widely employed in the classic MIMO communication system, which demands radio frequency (RF) chains comparable in number to the antennas \cite{7Jin2012, 8Jin2016}. Though multiple data streams could be transmitted simultaneously, the prohibitive energy consumption of these RF chains makes the fully digital precoding impractical for the mmWave IoT system \cite{7pi2011}. In the analog precoding architecture, all the antennas share a single RF chain \cite {8wang2009}. Using the phase shifters, analog precoding could obtain high precoding gains with low power consumption, but has to tolerate some performance loss. To circumvent the above problems, the hybrid precoding has been proposed, where a high-dimensional analog precoder is followed by a low-dimensional digital precoder\cite{9zhang2005, 10Venkateswaran2010, 11Ayach2013, 12Alkhateeb2013, Hanzo2018}. Between the analog and digital precoders, the number of the RF chains is much less than the number of antennas. The hybrid precoding could achieve similar performance as the digital precoding with much lower power consumption therefore it is more attractive to mmWave IoT systems.

There are many papers devote to the design of the hybrid precoding schemes\cite{11Ayach2013, 14Rusu2016, 15Singh2015, 13LDai2016, 19Chen2017}. The hybrid precoding architecture was first proposed for mmWave communications in \cite{11Ayach2013}. According to the sparse structure of the mmWave channel, an orthogonal matching pursuit (OMP) base scheme was proposed. This scheme firstly models the spectral efficiency optimization problem as a sparse reconstruction problem. Then, it selects the analog precoding vectors from the set of array response vectors and constructs the digital precoding matrix by the least square. Though the OMP scheme could achieve good spectral efficiency, it contains singular value decomposition (SVD) and inverse operations of high dimensional matrices, which lead to high computational complexity. Therefore, many recent hybrid precoding literatures focus on reducing the complexity of the hybrid precoding. In \cite{14Rusu2016}, the authors proposed four methods to achieve different tradeoffs between the performance and complexity for single user hybrid precoding. In \cite{15Singh2015, 13LDai2016}, the array-of-subarrays architecture was considered to reduce the computing complexity in which each RF chain is only collected with partial antennas. In \cite{19Chen2017}, the beamspace schemes were also proposed to obtain low-complexity hybrid precoding matrices, which transformed high-dimensional matrix operations into low-dimensional beamspace matrix operations. Moreover, the angular domain signal processing methods were also proposed for mmWave communications, which utilize array signal processing technologies to provide reliable design\cite{Wang2018, Zhao2018}.

To the best of our knowledge, all above methods were based on the clustered channel model and did not fully utilize the sparseness property in angular domain of the mmWave. According to \cite{ 21Samimi2014, 22Samimi2016, 23Rappaport2015}, the angles of the arrive/departure (AOAs/AODs) of the paths in the mmWave channel could be grouped in several separated spatial lobes (SLs). For paths in different spatial lobes, their AOAs/AODs are sufficiently separable, while the AOAs/AODs of the paths in one spatial lobe are relatively close. This sparseness property in the angular domain leads to the possibility to divide the mmWave channel approximately orthogonally, which could be utilized to reduce the complexity of the hybrid precoding and improve the system performance.

In this paper, we propose a low complexity hybrid precoding scheme and a diversity combining scheme for the mmWave IoT systems. By exploiting the sparseness property in angular domain of the mmWave, we firstly carry out a spatial lobes division (SLD) operation to group the total paths into several spatial lobes. SLD operation reconstructs the clustered mmWave channel into equivalent spatial lobes channel which consists of several approximately orthogonal sub-channels. Then, based on the SLD operation, we propose a low complexity hybrid precoding scheme, named HYP-SLD, which formulates the hybrid precoding design as a set of sparse reconstruction problems. For each sub-channel, the HYP-SLD scheme provides a decoupling solution to the design of the analog and digital precoding matrices. Finally, we further propose a maximum ratio combining based diversity combining scheme, named HYP-SLD-MRC. For data streams in each sub-channel, the HYP-SLD-MRC scheme adds data streams weighted by the corresponding signal-to-noise ratios (SNR) together for reducing the bit error rate (BER) of IoT.
The main contributions of this paper are summarized as follows.
\begin{itemize}
  \item We fully utilize the sparseness property in the angular domain of the mmWave to design the low complexity hybrid precoding scheme. The complexity of the proposed HYP-SLD scheme is proportional to the number of paths in one spatial lobe (sub-channel). Compared with the OMP scheme, the reduction of computational complexity is more than $99\%$ in a mmWave IoT system where the transmitter has 64 antennas and 16 RF chains, and the receiver has 32 antennas and 8 RF chains.
  \item Through a simple linear summation operation, the proposed HYP-SLD-MRC scheme maximizes the output SNR for each sub-channel. Therefore, the BER performance is greatly improved compared with the fully digital precoding scheme. Moreover, since the HYP-SLD-MRC scheme deals with each sub-channel rather than the total signals for the mmWave IoT systems, the multiplexing gains could also be obtained.
\end{itemize}

Simulation results demonstrate that the proposed HYP-SLD achieves near-optimal spectral efficiency and BER performances. Moreover, the proposed HYP-SLD-MRC scheme achieves great improvement in BER performance compared with the fully digital precoding scheme.

The rest of the paper is organized as follows. In Section II, the system model, channel model and the problem formulation are presented. The characteristics of spatial lobes, the equivalent spatial lobes channel and the low complexity hybrid precoding strategy are demonstrated in Section III. In Section IV, the diversity combining scheme is proposed. The simulation results are presented in Section V. Finally, we conclude this paper in Section VI.

We use the following notations in this paper: $a$ is a scalar, $\bf{a}$ is a vector, $\bf{A}$ is a matrix and ${\cal A}$ is a set. ${{\bf{A}}^{(i)}}$ is the $i_{th}$ column of $\bf{A}$ and ${\left\| {\bf{A}} \right\|_F}$ is the Frobenius norm of ${\bf{A}}$. ${{\bf{A}}^T},{{\bf{A}}^*},{{\bf{A}}^{ - 1}}$ denote the transpose, conjugate transpose and inverse of ${\bf{A}}$ respectively. ${\rm{diag}}({\bf{A}})$ is a vector that consists of diagonal elements of ${\bf{A}}$ and ${\rm{blkdiag}}({\bf{A}},{\bf{B}})$ is the block diagonal concatenation of ${\bf{A}}$ and ${\bf{B}}$. $[{\bf{A}}\left| {\bf{B}} \right.]$ is the horizontal concatenation. $\left| {\bf{a}} \right|$ is the modulus of $\bf{a}$. ${{\bf{I}}_N}$ denotes a
$N \times N$ identity matrix. ${\cal O}(N)$ means the order is $N$. ${\cal C}{\cal N}({\bf{a}},{\bf{A}})$ is a complex Gaussian vector with mean ${\bf{a}}$ and covariance matrix ${\bf{A}}$. $\mathbb{E}[{\bf{A}}]$ is the expectation of ${\bf{A}}$.
\section{System Model, Channel Model and Problem Formulation}
\subsection{System Model}
\begin{figure*}[t]
\centering
\includegraphics[scale=0.55]{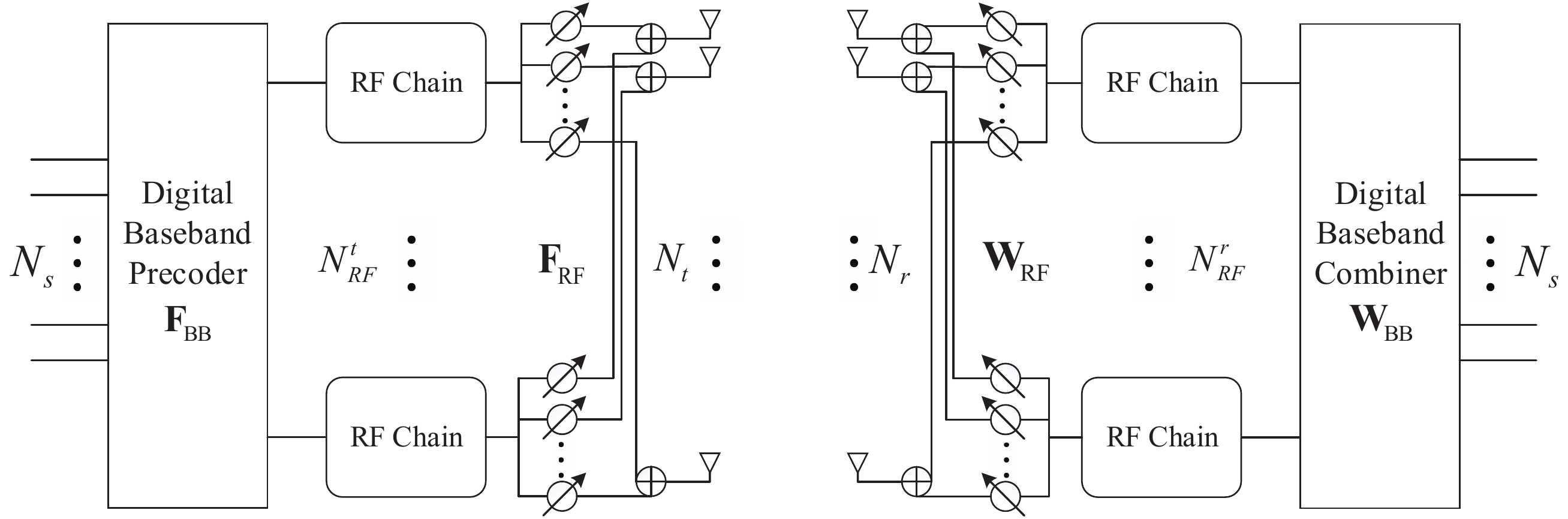}% 
\caption{Block diagram of the hybrid precoding structure in mmWave IoT systems.}
\label{fig1}
\end{figure*}
The hybrid precoding structure we consider in mmWave IoT systems is shown in Fig. \ref{fig1}. The transmitter and receiver of IoT devices are equipped with $N_{\rm{t}}$ and $N_{\rm{r}}$ antennas, respectively. The number of the RF chains at the transmitter and the receiver are respectively denoted as $N_{\rm{RF}}^{\rm{t}}$ and $N_{\rm{RF}}^{\rm{r}}$, which are subject to the constrains $N_{\rm{s}}\leq N_{\rm{RF}}^{\rm{t}}\leq N_{\rm{t}}$ and $N_{\rm{s}}\leq N_{\rm{RF}}^{\rm{r}}\leq N_{\rm{r}}$, where $N_{\rm{s}}$ denotes the number of the data streams.

%And the parameters satisfy $N_{\rm{s}}\leq N_{\rm{RF}}^{\rm{t}}\leq N_{\rm{t}}$ and $N_{\rm{s}}\leq N_{\rm{RF}}^{\rm{r}}\leq N_{\rm{r}}$.
At the transmitter, $N_{\rm RF}^{\rm t}\times N_{\rm s}$ baseband precoding matrix ${{\bf{F}}_{{\rm{BB}}}}$ followed by an $N_{\rm{t}}\times N_{\rm{RF}}^{\rm{t}}$ analog precoding matrix ${{\textbf{F}}_{{\rm{RF}}}}$ transforms $N_{\rm{s}}$ data streams to $N_{\rm{t}}$ antennas. Setting ${{\textbf{F}}_{{\rm{T}}}}={{\textbf{F}}_{{\rm{RF}}}}{{\textbf{F}}_{{\rm{BB}}}}$, the discrete-time transmitted signal vector could be written as
\begin{equation}\label{2.1}
\textbf{X}={{\textbf{F}}_{{\rm{T}}}}\textbf{s},
\end{equation}
where \textbf{s} is the $N_{\rm{s}}\times 1$ symbol vector with $\mathbb{E}[{\bf{s}}{{\bf{s}}^*}] = \frac{1}{{{N_{\rm s}}}}{{\bf{I}}_{{N_{\rm s}}}}$. In this system, ${{\textbf{F}}_{{\rm{RF}}}}$ is implemented by phase shifters, which has constant amplitude constraint ${\left( {{\bf{F}}_{{\rm{RF}}}^{(i)}{\bf{F}}_{{\rm{RF}}}^{(i)*}} \right)_{l,l}} = 1/{N_{\rm{t}}}$, where ${\left(  \cdot  \right)_{l,l}}$ denotes the ${l_{th}}$ diagonal element of a matrix. In addition, the total power constrain is enforced by $\left\| {{{\bf{F}}_{{\rm{RF}}}}{{\bf{F}}_{{\rm{BB}}}}} \right\|_F^2 = {N_{\rm{s}}}$.

We adopt a narrowband block-fading channel model as shown in \cite{11Ayach2013}, which yields the received signal as
\begin{equation}\label{2.2}
{\bf{r}} = \sqrt \rho  {\bf{H}}{{\bf{F}}_{\rm{T}}}{\bf{s}} + {\bf{n}},
\end{equation}
where ${\bf{H}}$ is the ${N_{\rm{r}}} \times {N_{\rm{t}}}$ mmWave channel matrix, $\rho$ is the average received power, and ${\bf{n}} \sim {\cal {CN}}(0,\sigma _n^{2})$ is the additive white Gaussian noise vector.

After being combined at the receiver, the received signal is
\begin{equation}\label{2.3}
{\bf{y}} = \sqrt \rho  {\bf{W}}_{\rm{T}}^*{\bf{H}}{{\bf{F}}_{\rm{T}}}{\bf{s}} + {\bf{W}}_{\rm{T}}^*{\bf{n}},
\end{equation}
where ${{\bf{W}}_{\rm{T}}} = {{\bf{W}}_{{\rm{RF}}}}{{\bf{W}}_{{\rm{BB}}}}$, ${{\bf{W}}_{{\rm{RF}}}}$ is the ${N_{\rm{r}}} \times N_{{\rm{RF}}}^{\rm{r}}$ RF combining matrix which should satisfy ${\left( {{\bf{W}}_{{\rm{RF}}}^{(i)}{\bf{W}}_{{\rm{RF}}}^{(i)*}} \right)_{l,l}} = 1/{N_{\rm{r}}}$ and ${{\bf{W}}_{{\rm{BB}}}}$ is the $N_{{\rm{RF}}}^{\rm{r}} \times {N_{\rm{s}}}$ baseband digital combining matrix.

\subsection{Channel Model}
The mmWave signals have higher free-space pathloss than lower frequency signals and are sensitive to blockages, which lead to limited spatial scattering. Therefore, the clustered channel model is usually used to represent the mmWave channel \cite{22Samimi2016}, which could be expressed as
%
%\[\sum\limits_{i = 1}^N i \]

\begin{equation}\label{2.4}
{\bf{H}} = \sqrt {\dfrac{{{N_{\rm{t}}}{N_{\rm{r}}}}}{{M{N}}}} \sum\limits_{m=1}^M {\sum\limits_{n=1}^{{N}} {{\alpha _{m,n}}} } {{\bf{a}}_{\rm{r}}}(\theta _{m,n}^{\rm r}){{\bf{a}}_{\rm{t}}}(\theta _{m,n}^{\rm t})^*,
\end{equation}
where $M$ is the number of clusters and each cluster contributes $N$ propagation paths, ${\alpha _{m,n}}$ denotes the complex gain of the $n_{th}$ path in the $m_{th}$ cluster, $\theta _{m,n}^r \in [0,2\pi ]$ and $\theta _{m,n}^t \in [0,2\pi ]$ are the AOA and AOD, respectively. By adopting uniform linear arrays (ULAs), the antenna array response vectors ${{\bf{a}}_{\rm{r}}}(\theta _{m,n}^{\rm r})$ and ${{\bf{a}}_{\rm{t}}}(\theta _{m,n}^{\rm t})$ at the transmitter and the receiver could be written as
\begin{equation}\label{2.5}
\begin{array}{l}
{{\bf{a}}_{\rm{t}}}(\theta _{m,n}^{\rm t}) = \dfrac{1}{{\sqrt {{N_{\rm t}}} }}\Big[1,{\kern 1pt} {e^{j(2\pi /\lambda )dsin(\theta _{m,n}^{\rm t}{\kern 1pt} )}},...{\kern 30pt}  \\
{\kern 80pt} ,{e^{j({N_{\rm t}} - 1)(2\pi /\lambda )dsin(\theta _{m,n}^{\rm t}{\kern 1pt} )}}{\Big]^T},
\end{array}
\end{equation}
and
\begin{equation}\label{2.6}
\begin{array}{l}
{{\bf{a}}_{\rm{r}}}(\theta _{m,n}^{\rm{r}}) = \dfrac{1}{{\sqrt {{N_{\rm{r}}}} }}\Big[1,{\kern 1pt} {e^{j(2\pi /\lambda )dsin(\theta _{m,n}^{\rm{r}}{\kern 1pt} )}},...{\kern 19pt} {\kern 1pt} {\kern 1pt} {\kern 1pt} {\kern 1pt} {\kern 1pt} {\kern 1pt} \\
 {\kern 78pt} ,{e^{j({N_{\rm{r}}} - 1)(2\pi /\lambda )dsin(\theta _{m,n}^{\rm{r}}{\kern 1pt} )}}{\Big]^T}{\kern 1pt} ,{\kern 1pt}
\end{array}
\end{equation}
respectively, where $\lambda$ is the wavelength of the signal, $d = \lambda /2$ denotes the aperture domain sample spacing. For convenient, we rewrite the channel in a more compact form as

\begin{equation}\label{2.7}
{\bf{H}}{\rm{ = }}{{\bf{A}}_{\rm{r}}}{\rm{diag}}({\boldsymbol{\alpha}}){{\bf{A}}_{\rm{t}}}^*,
\end{equation}
where ${\bf{\alpha }} = \sqrt {\frac{{{N_{\rm{t}}}{N_{\rm{r}}}}}{{MN}}} {[{\alpha _1},{\alpha _2},...,{\alpha _{{{MN}}}}]^T}$ contains the complex gains of all paths, and the matrices
\begin{equation}\label{2.8}
{{\bf{A}}_{\rm{r}}} = \big[{{\bf{a}}_{\rm{r}}}(\theta _{1,1}^{\rm{r}}),{{\bf{a}}_{\rm{r}}}(\theta _{1,2}^{\rm{r}}),...,{{\bf{a}}_{\rm{r}}}(\theta _{1,N}^{\rm{r}}),...,{{\bf{a}}_{\rm{r}}}(\theta _{M,N}^{\rm{r}})\big]
\end{equation}
and
\begin{equation}\label{2.9}
{{\bf{A}}_{\rm{t}}} = \big[{{\bf{a}}_{\rm{t}}}(\theta _{1,1}^{\rm{t}}),{{\bf{a}}_{\rm{t}}}(\theta _{1,2}^{\rm{t}}),...,{{\bf{a}}_{\rm{t}}}(\theta _{1,N}^{\rm{t}}),...,{{\bf{a}}_{\rm{t}}}(\theta _{M,N}^{\rm{t}})\big]
\end{equation}
contain the array response vectors. Inspired by (7), we could find that the number of the paths is the upper bound of the rank of the mmWave channel matrix.
%In this paper, we focus on the non-line-of- sight (NLOS) urban environments where the scattering is relevant rich. %From the spatial lobes perspective, we rewrite the millimeter wave channel (4).
%We divide the total paths into several spatial lobes and reconstruct the channel into equivalent spatial lobes channel which consist of several sub-channels. More details will be introduced in Section III B.

\subsection{Problem Formulation}
The target of designing the hybrid precoding matrices is to maximize the spectral efficiency of mmWave IoT systems achieved with Gaussian signalling over the mmWave channel \cite{27Goldsmith2003}, where the spectral efficiency is given by
\begin{equation}\label{2.10}
\begin{array}{l}
R = {\log _2}\Big (\Big | {{{\bf{I}}_{{N_{\rm{s}}}}}{\bf{ + }}\dfrac{\rho }{{{N_{\rm{s}}}}}{\bf{R}}_n^{ - 1}{\bf{W}}_{\mathop{\rm BB}\nolimits}^*{\bf{W}}_{\mathop{\rm RF}\nolimits}^*{\bf{H}}{{\bf{F}}_{{\mathop{\rm RF}\nolimits} }}{{\bf{F}}_{{\mathop{\rm BB}\nolimits} }}} \\
{\kern 91pt}   {{\kern 1pt}  {\bf{ \times F}}_{\mathop{\rm BB}\nolimits}^*{\bf{F}}_{\mathop{\rm RF}\nolimits}^*{{\bf{H}}^{\bf{*}}}{{\bf{W}}_{{\mathop{\rm RF}\nolimits} }}{{\bf{W}}_{{\mathop{\rm BB}\nolimits} }}} \Big |\Big ),
\end{array}
\end{equation}
where ${{\bf{R}}_{\rm{n}}} = \sigma _{\rm{n}}^2{\bf{W}}_{{\rm{BB}}}^ * {\bf{W}}_{{\rm{RF}}}^ * {{\bf{W}}_{{\rm{RF}}}}{{\bf{W}}_{{\rm{BB}}}}$ is the noise covariance matrix.
As shown in \cite{11Ayach2013}, the design of precoding matrices and combining matrices could be separated. The only difference is that the combining matrices do not have an extra power constraint. Therefore, we mainly focus on the design of the precoding matrices at the transmitter and the combining matrices at the receiver could be obtained similarly. The corresponding target of designing the precoding matrices could be simplified to maximize the mutual information, which is given by
%To simplify the design, the joint transmitter-receiver hybrid precoding design problem is decoupled and the goal of maximizing the spectral efficiency shown in (\ref{2.10}) could be replaced by maximizing the mutual information for the transmitter and receiver sides respectively.
\begin{equation}\label{2.11}
\begin{array}{l}
{{\cal I}_t}({\bf{F}}_{{\rm{RF}}},{\bf{F}}_{{\rm{BB}}}) = {\log_2}\Big (\Big |{{\bf{I}} + \dfrac{\rho }{{{N_{\rm{s}}}\sigma _n^2}}{\bf{H}}{{\bf{F}}_{{\rm{RF}}}}{{\bf{F}}_{{\rm{BB}}}}} {\kern 26pt} \\
{\kern 132pt}  {{\kern 1pt} {\kern 1pt}  \times {\bf{F}}_{{\rm{BB}}}^ * {\bf{F}}_{{\rm{RF}}}^ * {{\bf{H}}^ * }} \Big | \Big ).
\end{array}
\end{equation}
However, directly designing the precoding matrices to maximize (\ref{2.11}) is very non-trivial. Through mathematical derivation, the hybrid precoding design problem could be formulated as an equivalent sparse reconstruction problem which is aimed to minimize the Euclidean distance between the product of the analog and digital precoding matrices and the optimal unconstrained precoding matrix \cite{11Ayach2013}. The sparse reconstruction problem could be formulated as
\begin{equation}\label{2.13}
\begin{array}{l}
({\bf{F}}_{{\rm{RF}}}^{{\rm{opt}}},{\bf{F}}_{{\rm{BB}}}^{{\rm{opt}}}) = \mathop {{\rm{arg}}{\kern 1pt} {\kern 1pt} {\rm{min}}}\limits_{{{\bf{F}}_{{\rm{BB}}}},{{\bf{F}}_{{\rm{RF}}}}} {\left\| {{{\bf{F}}_{{\rm{opt}}}} - {{\bf{F}}_{{\rm{RF}}}}{{\bf{F}}_{{\rm{BB}}}}} \right\|_F},\\
 {\kern 83pt} {\rm{s}}{\rm{.t}}{\rm{.}}{\kern 7pt}  {{\bf{F}}_{{\rm{RF}}}} \in {{\cal F}_{{\rm{RF}}}},\\
 {\kern 97pt} \left\| {{{\bf{F}}_{{\rm{RF}}}}{{\bf{F}}_{{\rm{BB}}}}} \right\|_{_F}^2 = {N_s},
\end{array}
\end{equation}
where ${{\bf{F}}_{{\rm{opt}}}}$ is the optimal unconstrained precoding matrix which could be obtained from the SVD of the mmWave channel ${\bf{H}}$ and ${{\cal F}_{{\rm{RF}}}}$ is the set of the feasible RF precoders induced by the constant amplitude constraint. Note that, since the feasibility constraint on the RF precoding matrix is non-convex, it is very difficult to find a global optimal solution. In the design of our hybrid precoding scheme, we mainly exploit the sparseness property of the mmWave in the angular domain to find a low complexity near-optimal solution.

\section{Proposed Low Complexity Hybrid Precoding Algorithm Based on Spatial Lobes Division}
In this section, we firstly demonstrate the characteristic of spatial lobes of the mmWave channel. Then, we propose the SLD operation which reconstructs the mmWave channel into the equivalent spatial lobes channel. Based on the SLD operation, the low complexity hybrid precoding strategy is demonstrated in detail. Finally, we compare the complexity of the proposed hybrid precoding scheme with the OMP scheme.
%Finally, we introduce several extended situations.

\subsection{The Spatial Lobes Characteristics of mmWave}
\begin{figure}[htb]
\centering
\includegraphics[scale=0.15]{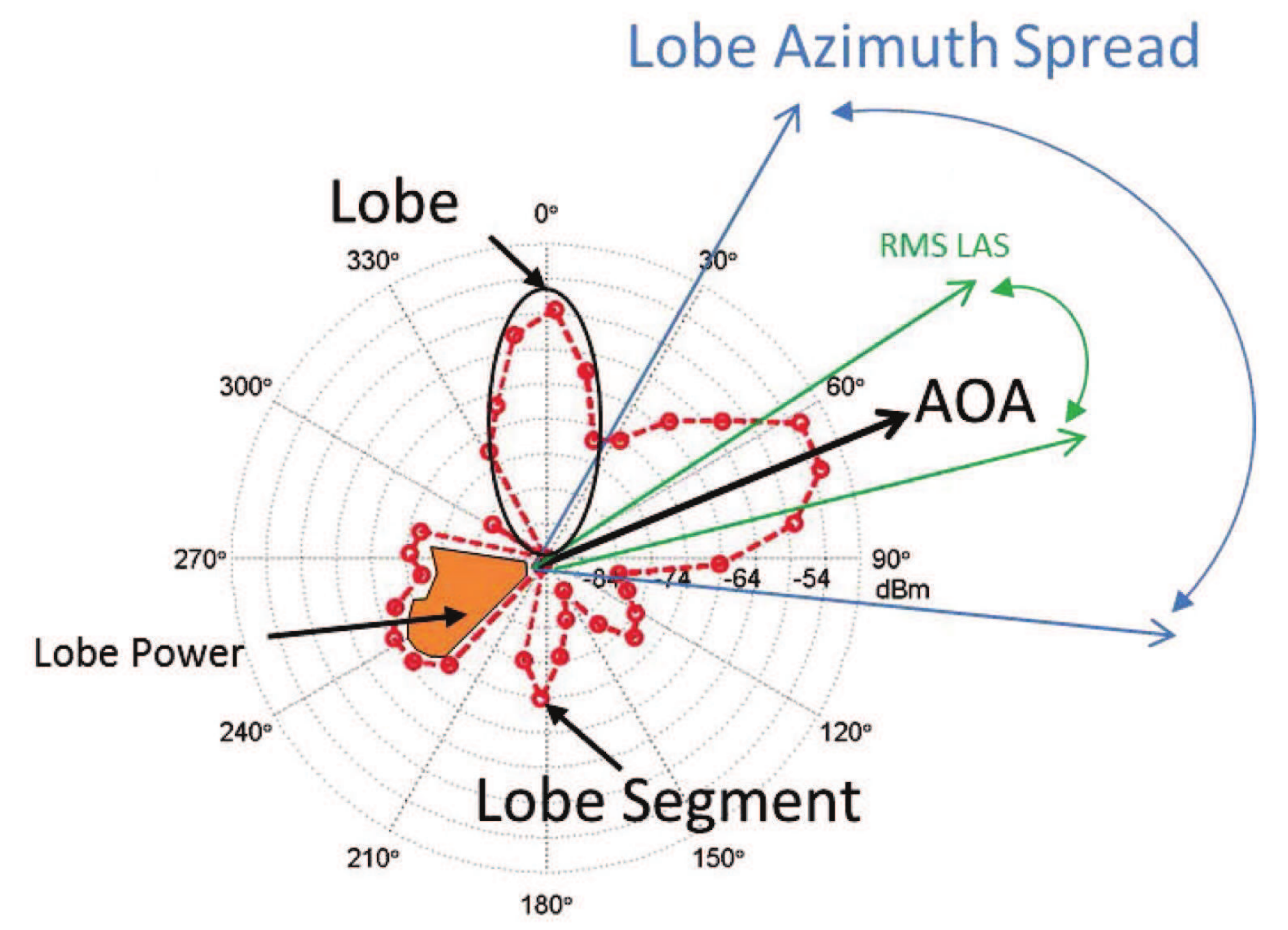}%fig2_polar_plot.jpg
\caption{The polar plot of mmWave channel measured in Manhattan at 28 GHz
\cite{21Samimi2014}.}
\label{fig2}
\end{figure}

%As we have introduced in Section I, the millimeter wave channel has also sparseness property in angle domain.
%{\color{Bittersweet}{ . }}
Recently, the mmWave channel was adequately measured by NYU WIRELESS which confirmed that the mmWave could be utilized in the 5G cellular networks \cite{21Samimi2014, 22Samimi2016, 23Rappaport2015}. The polar plot of 28 GHz mmWave channel \cite{21Samimi2014} is shown in Fig. \ref{fig2}. %In this measurement, they set the transmitter (TX) fixed at $269^\circ/-10^\circ$ and the receiver (RX) was omnidirectional.
At the receiver, there are five dominated spatial lobes with azimuth angle spreads, which confirms that the mmWave channels also have sparseness property in angular domain. In the traditional 3GPP and WINNER II channel models, which are widely used in LTE, the paths in one time cluster are assumed to arrive at a same angular spread. Whereas the measurement results by NYU WIRELESS indicate that there are some differences between the time cluster and the spatial lobe, which are summarized as follow.
\begin{itemize}
  \item The paths in one spatial lobe could come from more than one time cluster. Each spatial lobe represents a main AOA/AOD at which groups of multiple path components (MPCs) arrive/depart over a contiguous range of angles over several hundreds of nanoseconds \cite{23Rappaport2015}.
  \item A cluster may contain multipath components which travel close in time but arrive/depart from many angle lobe directions \cite{22Samimi2016}.
  \item The number of spatial lobes is independent of the number of time clusters \cite{22Samimi2016}.

\end{itemize}

The above differences indicate that the AOAs/AODs of paths in different time clusters may be close to each other. Therefore, we could not handle different time clusters separately. In contrast, the angles of paths in different spatial lobes are sufficiently separable, which prompts us to reconstruct the mmWave channel from the spatia lobes perspective and further to reduce the complexity of the hybrid precoding.

\begin{figure}[htb]
\centering
\includegraphics[scale=0.25]{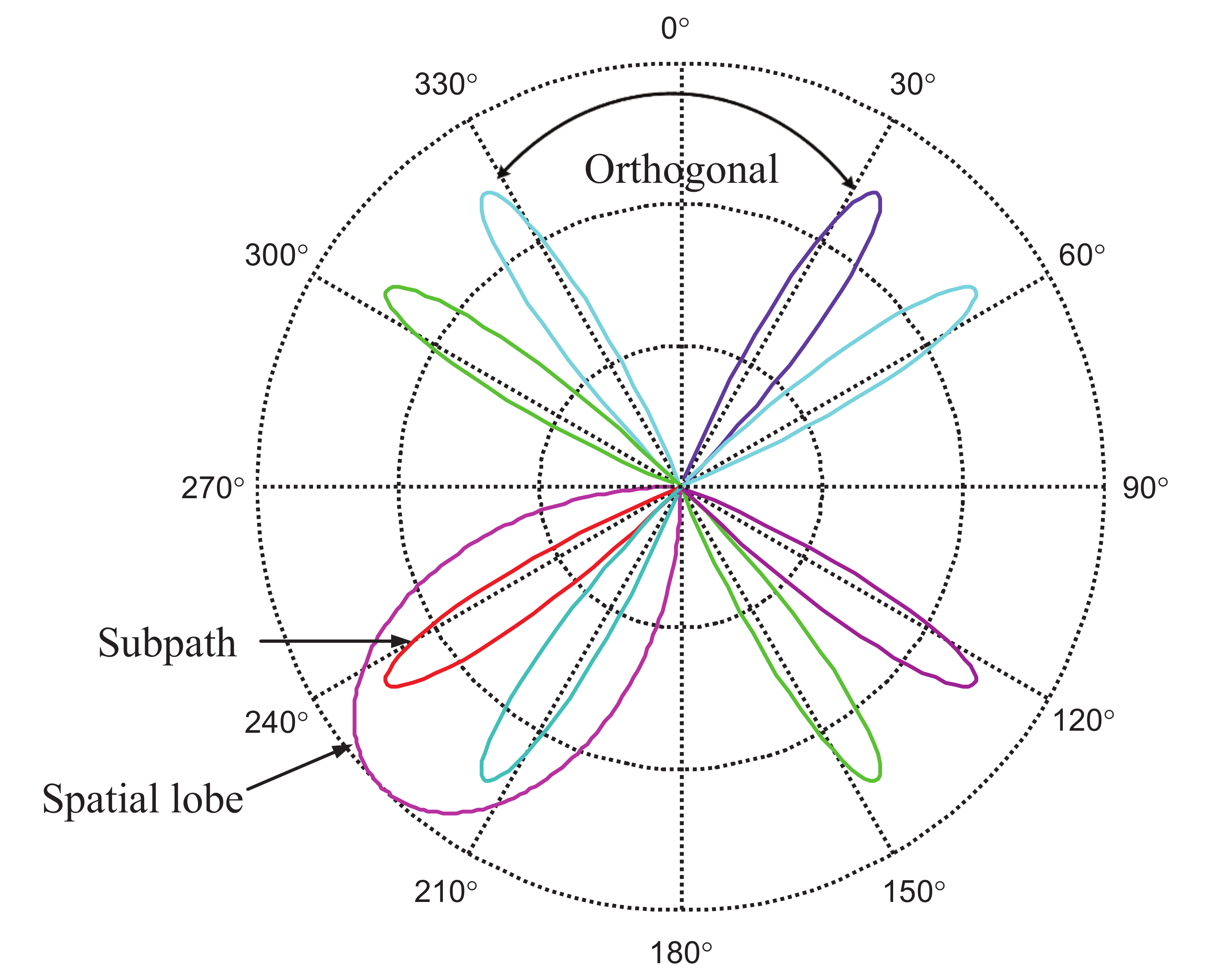}%polarplotbeam1.eps%58_polarplotbeam1.eps% angle-domain.eps %angle-domain2.eps
\caption{An example of the angular domain paths in the mmWave channel, where there are four spatial lobes and each spatial lobe contains two subpaths.}
\label{fig3}
\end{figure}

By exploiting the above characteristics of the spatial lobes and considering a relatively large number of antennas are usually employed in mmWave IoT systems, we make a reasonable assumption that the paths in different spatial lobes are approximately orthogonal since the AOAs/AODs of these paths are sufficiently separable. Therefore, the paths in the mmWave channel could be divided into several approximately orthogonal groups. An example of the angular domain distribution for the propagation paths considered in this paper is shown in Fig. \ref{fig3}, where there are four spatial lobes and each spatial lobe contains two subpaths.
%the  distributions of spatial lobe and subpaths

\subsection{SLD Operation and the Equivalent Spatial Lobes Channel}
As we could see, the clustered mmWave channel (4) is made of multiple propagation paths. Therefore, grouping the paths means dividing the channel. Based on the sparseness property of the mmWave in the angular domain, the SLD operation groups the total paths into several spatial lobes and reconstruct the mmWave channel (4) into the equivalent spatial lobe channel. Note that, the number of groups and the number of paths in each group are the number of spatial lobes and the number of sub-paths in each spatial lobe, respectively.

The equivalent spatial lobes channel could be written as
\begin{equation}\label{3.12}
\begin{array}{l}
{\bf{H}}_{\rm sl} = \sqrt {\dfrac{{{N_{\rm{t}}}{N_{\rm{r}}}}}{{P{Q}}}} \sum\limits_{p=1}^P {\sum\limits_{q=1}^{Q_p} {{\alpha _{p,q}}} } {{\bf{a}}_{\rm{r}}}({\theta _{p,q}^{\rm{r}})}{{\bf{a}}_{\rm{t}}}{(\theta _{p,q}^{\rm{t}})^*}\\
{\kern 15pt} = {{\bf{H}}_{\rm{1}}} + {{\bf{H}}_{\rm{2}}} + ... + {{\bf{H}}_{{P}}},
\end{array}
\end{equation}
where ${{\bf{H}}_{{i}}} = \sqrt {\dfrac{{{N_{\rm{t}}}{N_{\rm{r}}}}}{{P{Q}}}} \sum\limits_{q=1}^{Q_p} {{\alpha _{i,q}}} {{\bf{a}}_{\rm{r}}}({\theta _{i,q}^{\rm{r}})}{{\bf{a}}_{\rm{t}}}{(\theta _{i,q}^{\rm{t}})^*},{{i}}=1,2,...,{{P}}$  represents the $i_{th}$ sub-channel which contains the paths in the $i_{th}$ spatial lobe for both transmitter and receiver, $P$ is the number of spatial lobes, and $Q_p$ is the number of subpaths in the $p_{th}$ spatial lobe. According to [22, Section II-A] and [23, Section V-A], the maximum number of the spatial lobes is 5 and the mean angles of the spatial lobes are uniformly distributed between $0$ and $2\pi$, while the angles (AOAs/AODs) of the paths in one spatial lobe are randomly distributed within the range of the spatial lobe. Since the paths in different spatial lobes are approximately orthogonal, these sub-channels could be treated as approximately orthogonal to each other. The expression form of (13) is similar as the cluster channel model (4) and could be regarded as a reconstruction of (4). Therefore, we make ${\bf{H}}={\bf{H}}_{\rm sl}$ in the rest of the paper. We could also write (13) in a more compact expression as
\begin{equation}\label{2.7}
{\bf{H}}{\rm{ = }}{{\bf{A}}_{\rm{r}}}{\rm{diag}}({\boldsymbol{\alpha}}){{\bf{A}}_{\rm{t}}}^*,
\end{equation}
where ${{\bf{A}}_{\rm{r}}} = \big[{{\bf{a}}_{\rm{r}}}(\theta _{1,1}^{\rm{r}}),{{\bf{a}}_{\rm{r}}}(\theta _{1,2}^{\rm{r}}),...,{{\bf{a}}_{\rm{r}}}(\theta _{1,Q_1}^{\rm{r}}),...,{{\bf{a}}_{\rm{r}}}(\theta _{P,Q_P}^{\rm{r}})\big]$ and ${{\bf{A}}_{\rm{t}}} = \big[{{\bf{a}}_{\rm{t}}}(\theta _{1,1}^{\rm{t}}),{{\bf{a}}_{\rm{t}}}(\theta _{1,2}^{\rm{t}}),...,{{\bf{a}}_{\rm{t}}}(\theta _{1,Q_1}^{\rm{t}}),...,{{\bf{a}}_{\rm{t}}}(\theta _{P,Q_P}^{\rm{t}})\big] $. Accoding to the spatial lobe property, the above two antenna array response matrices could be divided into several parts as
 \begin{equation}\label{2.}
{{\bf{A}}_{\rm{t}}} = \big[{{\bf{A}}_{{\rm{t1}}}},{{\bf{A}}_{{\rm{t2}}}},...{{\bf{A}}_{{\rm{t}}P}}\big],
\end{equation}
 \begin{equation}\label{2.}
 {{\bf{A}}_{\rm{r}}} = \big[{{\bf{A}}_{{\rm{r1}}}},{{\bf{A}}_{{\rm{r2}}}},...{{\bf{A}}_{{\rm{r}}P}}\big],
 \end{equation}
 where
\begin{equation}\label{2.}
{{\bf{A}}_{{\rm{t}}i}} = \big[{{\bf{a}}_{\rm{t}}}(\theta _{i,1}^{\rm{t}}),{{\bf{a}}_{\rm{t}}}(\theta _{i,2}^{\rm{t}}),...,{{\bf{a}}_{\rm{t}}}(\theta _{i,Q_i}^{\rm{t}})\big],{\kern 1pt} {\kern 1pt} {\kern 1pt} {{i}}=1,2,...,{{P}}
\end{equation}
and
\begin{equation}\label{2.}
{{\bf{A}}_{{\rm{r}}i}} = \big[{{\bf{a}}_{\rm{r}}}(\theta _{i,1}^{\rm{r}}),{{\bf{a}}_{\rm{r}}}(\theta _{i,2}^{\rm{r}}),...,{{\bf{a}}_{\rm{r}}}(\theta _{i,Q_i}^{\rm{r}})\big],{\kern 1pt} {\kern 1pt} {\kern 1pt} {{i}}=1,2,...,{{P}}
\end{equation}
contain the antenna array response vectors of the $i_{th}$ sub-channel. According to (14)-(18), it could be obtained that $Q$ is the upper bound of the rank of the sub-channel. Therefore, we actually divide the mmWave channel into several low-rank approximately orthogonal sub-channels.

\subsection{Hybrid Precoding Based on Spatial Lobes Division}
In this subsection, we present a low complexity hybrid precoding scheme based on the SLD operation. In the design of the hybrid precoding, a principle of maximizing the usage of the channel is adopted, that is we keep the number of the data streams equal to the number of the total paths.
\begin{lemma}
(from \cite{Ayach2012}). {Each left and right singular vectors corresponding to non-zero eigenvalues of the matrix channel converge in chordal distance to the array response vectors when the number of the total paths ($L$) in the channel is much less than the number of antennas at both transmitter and receiver, i.e., $L = o({N_{\rm{t}}})$ and $L = o({N_{\rm{r}}})$.}
\end{lemma}

Lemma 1 demonstrates that the array response vectors are approximate orthogonal with each other and the channel representation (14) ``converges" to the SVD of {\bf{H}} for large number of the antennas.
\begin{corollary}
{When $L\ll min({N_{\rm t}, N_{\rm r}})$, the left and right singular matrices of the total channel matrix ${\bf{H}}$ consist of the left and right singular vectors from the sub-channels, respectively.}
\end{corollary}
\begin{IEEEproof}
For the $i_{th}$ sub-channel, we have
\begin{equation}\label{3.}
{\bf{H}}_{{i}} = {{\bf{A}}_{{\rm{r}}i}}{\rm{diag}}({\boldsymbol{\alpha}_i}){{\bf{A}}_{{\rm{t}}i}}^*={{\bf{U}}_{{i}}}{{\bf{\Sigma }}_{{i}}}{\bf{V}}_{{i}}^{\rm{*}}.
\end{equation}
%The optimal precoding matrix for $i_{th}$ sub-channel can be obtained by applying SVD
%\begin{equation}\label{3.}
%{\widetilde {\bf{H}}_{\rm{i}}}={{\bf{U}}_{\rm{i}}}{{\bf{\Sigma }}_{\rm{i}}}{\bf{V}}_{\rm{i}}^{\rm{*}}. i=1,2,...,P
%\end{equation}
According to (15) and (16), the total array response matrices consist of the array response matrices for each sub-channel. In the meantime, according to Lemma 1, we could conclude that left and right singular vectors corresponding to the $Q$ largest singular values in ${{\bf{U}}_{{i}}}$ and ${\bf{V}}_{{i}}^{{*}}$ converge in chordal distance to the responding array response vectors in ${{\bf{A}}_{{\rm{r}}i}}$ and ${{\bf{A}}_{{\rm{t}}i}}$, respectively. %Therefore, we prove the corollary.
\end{IEEEproof}

Since the left and right singular matrices are the optimal unconstrained fully digital precoding matrices, Corollary 1 indicates that, the design of the precoding matrix for the total channel matrix could be divided into the precoding design for each sub-channel when $L\ll min({N_{\rm t}, N_{\rm r}})$. Note that the angles of antenna response vectors for different spatial lobes are sufficiently separable, which makes the ``inter-lobes" interference very small.
Thus, even the antenna response vectors are not orthogonal to each other for the antennas array of practical size, we could still divide the total hybrid precoding problem into several subproblems, each of which is only designed for one sub-channel.%, since the angles of antenna response vectors for different spatial lobes are sufficiently separable

Therefore, for the $i_{th}$ spatial lobe or sub-channel, the optimization problem could be formulated as
%\begin{small}
\begin{equation}\label{3.15}
\begin{array}{l}
({\bf{F}}_{{\rm{RF}}_i}^{{\rm{opt}}},{\bf{F}}_{{\rm{BB}}_i}^{{\rm{opt}}}) = \mathop {{\rm{arg}}{\kern 1pt} {\kern 1pt} {\rm{min}}}\limits_{{{\bf{F}}_{{\rm{BB}}_i}},{{\bf{F}}_{{\rm{RF}}_i}}} {\left\| {{{\bf{F}}_{{\rm{opt}}_i}} - {{\bf{F}}_{{\rm{RF}}_i}}{{\bf{F}}_{{\rm{BB}}_i}}} \right\|_F},{\kern 16pt} \\
{\kern 71pt} {\rm{s}}{\rm{.t}}{\rm{.}}{\kern 1pt} {\kern 1pt} {\kern 1pt} {\kern 1pt} {\kern 1pt} {\kern 1pt} {\kern 1pt} {{\bf{F}}_{{\rm{RF}}_i}} \in {{\cal F}_{{\rm{RF}}}},\\
{\kern 86pt}\left\| {{{\bf{F}}_{{\rm{RF}}_i}}{{\bf{F}}_{{\rm{BB}}_i}}} \right\|_{F}^2 = {N_{\rm s}{{{A_i}}}\bigg /{{\sum\limits_{i = 1}^P {{A_i}}}}},
\end{array}
\end{equation}
%\end{small}
where ${{\bf{F}}_{{\rm{opt}}_i}}={\bf{V}}_{{i}}(:,1:Q_i)$ is the optimal reference precoding matrix, ${{\bf{F}}_{{\rm{RF}}_i}}$ and ${{\bf{F}}_{{\rm{BB}}_i}}$ are the analog precoding matrix and digital precoding matrix for the $i_{th}$ sub-channel, respectively. ${{\cal F}_{{\rm{RF}}}}=\mathop  \bigcup \limits_{{{i}}=1,2,...,{{P}}} {{\cal F}_{{\rm{RF}}_i}}$ is the set of the feasible RF precoders and ${{\cal F}_{{\rm{RF}}_i}}$ is the feasible set of RF precoder for the $i_{th}$ sub-channel. $A_i$ is the total power in the $i_{th}$ spatial lobe and we assume equal power distribution in this paper.

Note that, since the number of paths $Q_i$ in each sub-channel is very small, i.e., $Q_i \ll \min ({N_{\rm t}},{N_{\rm r}})$, the sub-channel could be considered to be in a very poor scattering environment.
Inspired by Lemma 1, in the proposed hybrid precoding scheme, we set the antenna array response matrices ${{\bf{A}}_{{\rm{t}}i}}$ and ${{\bf{A}}_{{\rm{r}}i}}$ as the reference matrices ${{\bf{F}}_{{\rm{res}}}}$, rather than the fully digital precoding matrix obtained by high-dimensional SVD, for the $i_{th}$ sub-channel. However, for arrays of practical sizes, only setting ${{\bf{A}}_{{\rm{t}}i}}$ and ${{\bf{A}}_{{\rm{r}}i}}$ as the reference precoding matrices may cause many performance losses. Therefore, we further perform a digital precoding design at the baseband. In summary, in our hybrid precoding scheme, we decouple the solution of the optimization problem (20) into analog and digital phases, where the target of the analog precoding is to find the constant amplitude vectors which are closest to each entry of the ${{\bf{F}}_{{\rm{res}}}}$ in the $l_2$ norm sense, and the digital precoding is aimed to remove the interference and perform power allocation to these vectors. We demonstrate the design of the analog and digital precoding matrices in detail as follows.
\begin{lemma}
{For the selected vectors of different sub-channels, there is no overlap between the corresponding feasible sets ${{\cal F}_{{\rm{RF}}_i}}, {{i}}=1,2,...,{{P}}$.}
\end{lemma}
\begin{IEEEproof}
Define the beam coverage of the $i_{th}$ spatial lobe as
\begin{equation}\label{3.}
\small
{\bf{\cal{CV}}}({SL}_{{i}}) = \mathop  \bigcup \limits_{{{j}} = 1,2,...,Q} {\cal{CV}}\big(a(\theta _{{{i,j}}}^{\rm{t}})\big){\kern 1pt} {\kern 1pt} ,{\kern 1pt} {\kern 1pt} {\kern 1pt} {\kern 1pt} {{i}}=1,2,...,{{P}},
\end{equation}
where ${\cal{CV}}\big(a(\theta _{{{i,j}}}^{\rm{t}})\big)$ is the beam coverage of the steering vectors in the $i_{th}$ spatial lobe (sub-channel). For the ULA considered in this paper, the half-power beam width of the array is approximately equal to $102^\circ/N$ \cite{28Tse}, where $N$ is the number of antennas, i.e., $length({\cal{CV}}\big(a(\theta _{{{i,j}}}^{\rm{t}})\big))=102^\circ/N$ ($length({\cal{CV}}(.))$ represents the beam width of ${\cal{CV}}(.)$). Therefore, we have
\begin{equation}\label{3.}
\frac{102^\circ}{N}\leq length({\cal{CV}}(S{L_{{i}}})) \leq Q_i\frac{102^\circ}{N}.
\end{equation}
%\end{small}
As shown in Fig. \ref{fig3}, the AOAs and AODs in different spatial lobes are sufficiently separable. When the angle interval $\bigtriangleup_\theta $ between the mean angles of different spatial lobes satisfies
\begin{equation}\label{3.}
\bigtriangleup_\theta > Q_i\frac{102^\circ}{N},
\end{equation}
there will be no overlap between different spatial lobes. Moreover, since the angles of ${\bf{\cal{CV}}}({{\cal F}_{{\rm{RF}}_i}})$ could not exceed the range of the $i_{th}$ spatial lobe,
we have
\begin{equation}\label{3.}
{\bf{\cal{CV}}}({{\cal F}_{{\rm{RF}}_i}})\subseteq {\bf{\cal{CV}}}({SL}_{{i}}).
\end{equation}
Therefore, there will be no overlap between the different feasible sets.
%That is the corresponding feasible sets for different sub-channel have no overlap.
%The azimuth AOAs and AODs are drawn independently from a continuous distribution in each spatial lobe. Moreover, the paths in different sub-channel are belong to different spatial lobes, which makes their AODs and AOAs sufficiently separable. Therefore, the selected vectors for $i_{th}$ sub-channel will not be selected for $j_{th} (j\neq i$) sub-channel and the vectors belong to different spatial lobes will cause little interference to each other.
\end{IEEEproof}

%demonstrates the orthogonality between these spatial lobe sub-channels and
Lemma 2 indicates that the selected vectors for different sub-channel cause small interference with each other and the total feasible set ${{\cal F}_{{\rm{RF}}}}$ could be simply divided into $P$ parts for each sub-channel to select the vectors in parallel. Actually, this is why we assume that the paths in different spatial lobes are approximately orthogonal with each other in section III.A. Therefore, (20) could be simplified as
\begin{equation}\label{3.15}
%\begin{array}{l}
%({\bf{F}}_{{\rm{RFi}}}^{{\rm{opt}}},{\bf{F}}_{{\rm{BBi}}}^{{\rm{opt}}}) = \mathop {{\rm{arg}}{\kern 1pt} {\kern 1pt} {\rm{min}}}\limits_{{{\bf{F}}_{{\rm{BBi}}}},{{\bf{F}}_{{\rm{RFi}}}}} {\left\| {{{\bf{F}}_{{\rm{opti}}}} - {{\bf{F}}_{{\rm{RFi}}}}{{\bf{F}}_{{\rm{BBi}}}}} \right\|_F},{\kern 16pt} \\
% {\kern 91pt} {\rm{s}}{\rm{.t}}{\rm{.}}{\kern 1pt} {\kern 1pt} {\kern 1pt} {\kern 1pt} {\kern 1pt} {\kern 1pt} {\kern 1pt} {{\bf{F}}_{{\rm{RFi}}}} \in {{\cal F}_{{\rm{RFi}}}},\\
%{\kern 106pt}\left\| {{{\bf{F}}_{{\rm{RFi}}}}{{\bf{F}}_{{\rm{BBi}}}}} \right\|_{_F}^2 = {N_s}\dfrac{{{A_i}}}{{\sum\limits_{i = 1}^P {{A_i}} }}.
%\end{array}
\begin{array}{l}
({\bf{F}}_{{\rm{RF}}_i}^{{\rm{opt}}},{\bf{F}}_{{\rm{BB}}_i}^{{\rm{opt}}}) = \mathop {{\rm{arg}}{\kern 1pt} {\kern 1pt} {\rm{min}}}\limits_{{{\bf{F}}_{{\rm{BB}}_i}},{{\bf{F}}_{{\rm{RF}}_i}}} {\left\| {{{\bf{F}}_{{\rm{opt}}_i}} - {{\bf{F}}_{{\rm{RF}}_i}}{{\bf{F}}_{{\rm{BB}}_i}}} \right\|_F},{\kern 16pt} \\
{\kern 71pt} {\rm{s}}{\rm{.t}}{\rm{.}}{\kern 1pt} {\kern 1pt} {\kern 1pt} {\kern 1pt} {\kern 1pt} {\kern 1pt} {\kern 1pt} {{\bf{F}}_{{\rm{RF}}_i}} \in {{\cal F}_{{\rm{RF}}_i}},\\
{\kern 86pt}\left\| {{{\bf{F}}_{{\rm{RF}}_i}}{{\bf{F}}_{{\rm{BB}}_i}}} \right\|_{F}^2 = {N_{\rm s}{{{A_i}}}\bigg /{{\sum\limits_{i = 1}^P {{A_i}}}}},
\end{array}
\end{equation}

In the analog precoding phase, to make the precoding scheme more practical for the limited feedback system, feasible sets are quantized with limited $b$ bits \cite{29Lin2017}. The quantized candidate matrix for the transmitter is
\begin{equation}\label{3.}
{\bf{A}}_{\rm{t}}^{{\rm{quant}}} = \Big[a_{\rm{t}}^{{\rm{quant}}}({\theta _1}),a_{\rm{t}}^{{\rm{quant}}}({\theta _2}),...,a_{\rm{t}}^{{\rm{quant}}}({\theta _{{2^b}}})\Big],
\end{equation}

where the entries of ${\bf{A}}_{\rm{t}}^{{\rm{quant}}}$ are
\begin{equation}\label{3.}
\small
a_{\rm{t}}^{{\rm{quant}}}({\theta _i}) = \dfrac{1}{{\sqrt {{N_{\rm t}}} }}\Big[1,{e^{j\pi {\rm sin}(\frac{{2\pi (i-1)}}{{{2^b}}})}},...,{e^{j({N_{\rm t}} - 1)\pi {\rm sin}(\frac{{2\pi (i-1)}}{{{2^b}}})}} \Big]^T.
%a_{\rm{t}}^{{\rm{quant}}}({\theta _i}) = \dfrac{1}{{\sqrt {{N_{\rm t}}} }}\Big[1,{e^{\frac{{j2\pi (i-1)}}{{{2^b}}}}},...,{e^{\frac{{j2\pi ({N_{\rm t}} - 1)(i - 1)}}{{{2^b}}}}}\Big]^T.
\end{equation}

Note that, the quantization candidate matrix divides the angular domain space into $2^b$ parts uniformity, and could be further divided into $P$ spatial lobes parts as
\begin{equation}\label{3.}
{{\bf{A}}_{\rm{t}}^{{\rm{quant}}}} = \Big[{\bf{A}}_{\rm{t1}}^{{\rm{quant}}},{\bf{A}}_{\rm{t2}}^{{\rm{quant}}},...,{\bf{A}}_{{\rm{t}}P}^{{\rm{quant}}}\Big].
\end{equation}

%The selection of quantization bit is related to the number of antennas.
%Generally, we should at least make the quantization bit satisfy $2^b\geq max({N_{\rm t}, N_{\rm r}})$ to guarantee good performance. However, the complexity increases exponentially with the quantization bit. Therefore, selecting a reasonable quantization bit is important, and the corresponding simulation will be presented in Section V.

Given the quantization matrices, the remaining operations in the analog precoding phase are to find the vectors form the quantization matrices which are closest to each entry of the ${{\bf{F}}_{{\rm{res}}}}$ in the $l_2$ norm sense. This is equivalent to find the vectors along which the reference matrix ${{\bf{F}}_{{\rm{res}}}}$ has the maximum projection. We only introduce the analog precoding design at the transmitter, while the analog precoding matrix at the receiver could be obtained in the same way. The correlation matrix is
\begin{equation}\label{3.}
{\bf{\Psi}}  = {\bf{A}}_{{\rm{t}}i}^{{\rm{quant*}}}{{\bf{F}}_{{\rm{res}}}}.
\end{equation}
The power distributed in each direction could be calculated as
\begin{equation}\label{3.}
{\bf{k}} = {\rm{diag}}({\bf{\Psi }}{{\bf{\Psi }}^*}).
\end{equation}
Then, we select the position indexes of the $Q_i$  largest values in ${\bf{k}}$, and obtain the corresponding vectors from the quantization matrices. Once a vector has been selected, the value of the corresponding location in ${\bf{k}}$ is set to be zero to eliminate the effects of the vector. According to (29) and (30), the analog precoding matrix for the $i_{th}$ sub-channel could be obtained as
\begin{equation}\label{3.}
{{\bf{F}}_{{\rm{RF}}_i}} = [{{\bf{F}}_{{\rm{RF}}_{i1}}},{{\bf{F}}_{{\rm{RF}}_{i2}}},...,{{\bf{F}}_{{\rm{RF}}_{iQ_i}}}], {{i}}=1,2,...,{{P}},
\end{equation}
%\begin{equation}\label{3.}
%{{\bf{W}}_{\rm{RF}_i}} = [{{\bf{W}}_{\rm{RF}_{i1}}},{{\bf{W}}_{\rm{RF}_{i2}}},...,{{\bf{W}}_{\rm{RF}_{iQ}}}],
%\end{equation}
where ${{\bf{F}}_{{\rm{RF}}_{ij}}}$ represents the selected vectors steering at the $j_{th}$ paths in the $i_{th}$ sub-channel.
After the analog precoding matrices for all sub-channels are obtained, the final analog precoding matrix at the transmitter could be determined by
\begin{equation}\label{3.}
{{\bf{F}}_{{\rm{RF}}}} = [{{\bf{F}}_{\rm{RF}_1}},{{\bf{F}}_{\rm{RF}_2}},...,{{\bf{F}}_{{\rm{RF}}_{P}}}].
 \end{equation}
Note that, since the selected vectors ${\bf{F}}_{{\rm{RF}}_{ij}}$ are not orthogonalized, we only need to find these vectors to steer at the paths and leave the orthogonalization process to the digital precoding phase.
In the similar way, the analog precoding matrix at the receiver could be obtained as
\begin{equation}\label{3.}
{{\bf{W}}_{{\rm{RF}}}} = [{{\bf{W}}_{\rm{RF}_1}},{{\bf{W}}_{\rm{RF}_2}},...,{{\bf{W}}_{{\rm{RF}}_{P}}}].
\end{equation}

After the analog precoding phase, we obtain the effective low-dimensional channel as
\begin{equation}\label{3.22}
\begin{array}{l}
{{\bf{H}}_{{\rm{eq}}}} = {\bf{W}}_{{{\rm{RF}}}}^ * {\bf{H}}{{\bf{F}}_{{\rm{RF}}}}\\
 = {\left[{{\bf{W}}_{{\rm{R}}{{\rm{F}}_1}}},{{\bf{W}}_{{\rm{R}}{{\rm{F}}_{\rm{2}}}}},...,{{\bf{W}}_{{\rm{R}}{{\rm{F}}_{{P}}}}}\right]^ * }{\bf{H}} \left[{{\bf{F}}_{{\rm{R}}{{\rm{F}}_{\rm{1}}}}},{{\bf{F}}_{{\rm{R}}{{\rm{F}}_{\rm{2}}}}},...,{{\bf{F}}_{{\rm{R}}{{\rm{F}}_{{P}}}}}\right] \\
 = \left[ \begin{array}{l}
{\bf{W}}_{{{\rm{R}}{{\rm{F}}_1}}}^*{\bf{H}}{{\bf{F}}_{{\rm{R}}{{\rm{F}}_{\rm{1}}}}},{\kern 1pt} {\kern 1pt} {\kern 1pt} {\kern 1pt} {\bf{W}}_{{{\rm{R}}{{\rm{F}}_1}}}^*{\bf{H}}{{\bf{F}}_{{\rm{R}}{{\rm{F}}_{\rm{2}}}}}{\kern 1pt} {\kern 1pt} , \ldots ,{\bf{W}}_{{{\rm{R}}{{\rm{F}}_1}}}^*{\bf{H}}{{\bf{F}}_{{\rm{R}}{{\rm{F}}_{{P}}}}}\\
{\bf{W}}_{{{\rm{R}}{{\rm{F}}_2}}}^*{\bf{H}}{{\bf{F}}_{{\rm{R}}{{\rm{F}}_{\rm{1}}}}},{\kern 1pt} {\kern 1pt} {\kern 1pt} {\kern 1pt} {\bf{W}}_{{{\rm{R}}{{\rm{F}}_2}}}^*{\bf{H}}{{\bf{F}}_{{\rm{R}}{{\rm{F}}_2}}}{\kern 1pt} {\kern 1pt} , \ldots ,{\bf{W}}_{{{\rm{R}}{{\rm{F}}_2}}}^*{\bf{H}}{{\bf{F}}_{{\rm{R}}{{\rm{F}}_{{P}}}}}\\
 {\kern 30pt}  \vdots {\kern 64pt}  \vdots {\kern 30pt}  \ddots {\kern 34pt}   \vdots \\
{\bf{W}}_{{{\rm{R}}{{\rm{F}}_{{P}}}}}^*{\bf{H}}{{\bf{F}}_{{\rm{R}}{{\rm{F}}_{\rm{1}}}}},{\kern 1pt} {\kern 1pt} {\kern 1pt}  {\bf{W}}_{{{\rm{R}}{{\rm{F}}_{{P}}}}}^*{\bf{H}}{{\bf{F}}_{{\rm{R}}{{\rm{F}}_2}}}{\kern 1pt} , \ldots  ,{\bf{W}}_{{{\rm{R}}{{\rm{F}}_{{P}}}}}^*{\bf{H}}{{\bf{F}}_{{\rm{R}}{{\rm{F}}_{{P}}}}}
\end{array} \right]{\kern 1pt} \\
 = \left[ \begin{array}{l}
{\widetilde {\bf{H}}_{{\rm{11}}}},{\widetilde {\bf{H}}_{{\rm{12}}}}, \ldots ,{\widetilde {\bf{H}}_{{{1P}}}}\\
{\widetilde {\bf{H}}_{{\rm{21}}}},{\widetilde {\bf{H}}_{{\rm{22}}}}, \ldots ,{\widetilde {\bf{H}}_{{{2P}}}}\\
 {\kern 6pt}  \vdots {\kern 21pt}  \vdots {\kern 13pt}   \ddots {\kern 10pt}   \vdots \\
{\widetilde {\bf{H}}_{{\rm{P1}}}},{\widetilde {\bf{H}}_{{\rm{P2}}}}, \ldots ,{\widetilde {\bf{H}}_{{{PP}}}}
\end{array} \right],
\end{array}
\end{equation}
where ${\widetilde{\bf{H}}_{{ij}}} \in \mathbb{C}{^{Q_i \times Q_j}}$ are called as effective sub-channels.
%For the sub-channels in same column of the equivalent channel, they have the same spatial lobe in the transmitter, and for the sub-channels in same column of the equivalent channel, they have the same spatial lobe in the receiver. And considering that the angles of different spatial lobes are sufficiently separable so the sub-channels can be regarded as ``orthogonal'' to each other.
According to (14), we could find that ${{\bf{W}}_{{\rm{RF}}_i}}$ and ${{\bf{F}}_{{\rm{RF}}_i}}$ are one by one correspondence for the $i_{th}$ sub-channel. Therefore, only the diagonal effective sub-channels make sense, and (34) could be rewritten as
\begin{equation}\label{4.}
{\widetilde {\bf{H}}_{{\rm{eq}}}}= \left[ \begin{array}{l}
{\widetilde {\bf{H}}_{{\rm{11}}}}\\
{\kern 19pt}  \ddots  \\
{\kern 33pt} {\widetilde {\bf{H}}_{{{PP}}}}
\end{array} \right],
\end{equation}
where ${\widetilde {\bf{H}}_{{{ii}}}}$ contains the paths whose AODs and AOAs belong to the $i_{th}$ spatial lobe. %Therefore, these effective sub-channels can be treated as orthogonal to each other.
\begin{algorithm}[t]
    \renewcommand{\algorithmicrequire}{\textbf{Input:}}
	\renewcommand{\algorithmicensure}{\textbf{Output:}}
	\caption{Hybrid Precoding Based on Spatial Lobes Division (HYP-SLD)}
    \label{alg:1}
	\begin{algorithmic}[1]
        \REQUIRE %input
        $ {{\bf{A}}_{\rm{t}}}$, ${{\bf{A}}_{\rm{r}}}$, ${{\bf{A}}_{\rm t}^{{\rm{quant}}}}$, ${{\bf{A}}_{\rm r}^{{\rm{quant}}}}$
		\ENSURE ${{\bf F}_{{\rm{RF}}}}$, ${{\bf F}_{{\rm{BB}}}},{{\bf W}_{{\rm{RF}}}}$, ${{\bf W}_{{\rm{BB}}}}$
		\FOR {$i\leq P$}
        %\STATE ${\bf{F}}_{{\rm{res}}}={{\bf{A}}_{\rm{t}}}(:,(p-1)\ast Q +1:p\ast Q)$
        \STATE ${\bf{F}}_{{\rm{res}}}={{\bf{A}}_{{\rm{ti}}}}$
        \STATE ${\bf{\Psi}}  = {\bf{A}}_{{\rm{t}}i}^{{\rm{quant*}}}{{\bf{F}}_{{\rm{res}}}}$
		\FOR {$j\leq Q_i$}
		%\STATE $A_{\rm{t}}^{{\rm{quant}}}$
        \STATE $k \leftarrow {\rm{argma}}{{\rm{x}}_{l = 1,...,{\rm{Q}}}}{\rm{diag}}({\bf{\Psi }}{{\bf{\Psi }}^*})$
        \STATE ${{\bf{F}}_{{\rm RF}_{i}}} = \Big[{{\bf{F}}_{{\rm{RF}}_i}}\left| {{\bf{A}}_{{\rm{t}}i}^{{\rm{quant}}(k)}} \right.\Big]$
        \STATE Eliminate the effect of the selected vector ${\rm{diag}}({\bf{\Psi }}{{\bf{\Psi }}^*})(l)=0$
		\ENDFOR
        \ENDFOR
        \STATE${{\bf{F}}_{{\rm{RF}}}} \leftarrow [{{\bf{F}}_{{\rm{RF_1}}}},{{\bf{F}}_{{\rm{RF_2}}}},...,{{\bf{F}}_{{\rm{RF}}_P}}]{\kern 1pt} {\kern 1pt}$
        \STATE We could obtain ${{\bf{W}}_{{\rm{RF}}}}$ in the same way\\
        \STATE ${{\bf{W}}_{{\rm{RF}}}} = [{{\bf{W}}_{{\rm{RF_1}}}},{{\bf{W}}_{{\rm{RF_2}}}},...,{{\bf{W}}_{{\rm{RF}}_P}}]$
        \STATE ${{\bf{H}}_{{\rm{eq}}}} = {\bf{W}}_{{\rm{RF}}}^{\rm{*}}{\bf{H}}{{\bf{F}}_{{\rm{RF}}}}$
		\FOR {$i\leq P$}
        \STATE Compute the SVD of ${\bf{W}}_{{\rm{RF}}_i}^{\rm{*}}{\bf{H}}{{\bf{F}}_{{\rm{RF}}_i}}$ from ${{\bf{H}}_{{\rm{eq}}}}$
        \STATE ${\bf{W}}_{{\rm{RF}}_i}^{\rm{*}}{\bf{H}}{{\bf{F}}_{{\rm{RF}}_i}} = {{\bf{U}}_{{ii}}}{{\bf{\Sigma }}_{{ii}}}{\bf{V}}_{{ii}}^*$
        \STATE ${{\bf{F}}_{{\rm{BB}}_i}} = {{\bf{V}}_{{ii}}}$,${{\bf{W}}_{{\rm{BB}}_i}} = {{\bf{U}}_{{ii}}}$
        \ENDFOR
        \STATE ${{\bf{F}}_{{\rm{BB}}}} = {\rm{blkdiag}}({{\bf{F}}_{{\rm{BB_1}}}},{{\bf{F}}_{{\rm{BB_2}}}},...,{{\bf{F}}_{{\rm{BB}}_P}})$
        \STATE ${{\bf{W}}_{{\rm{BB}}}} = {\rm{blkdiag}}({{\bf{W}}_{{\rm{BB_1}}}},{{\bf{W}}_{{\rm{BB_2}}}},...,{{\bf{W}}_{{\rm{BB}}_P}})$
        \STATE ${{\bf{F}}_{{\rm{BB}}}} = \sqrt {N_{\rm s}} \frac{{{{\bf{F}}_{{\rm{BB}}}}}}{{{{\left\| {{{\bf{F}}_{{\rm{RF}}}}{{\bf{F}}_{{\rm{BB}}}}} \right\|}_F}}}$%${{\bf{W}}_{{\rm{BB}}}} = \sqrt {{N_s}} \frac{{{{\bf{W}}_{{\rm{BB}}}}}}{{{{\left\| {{{\bf{W}}_{{\rm{RF}}}}{{\bf{W}}_{{\rm{BB}}}}} \right\|}_F}}}$
    \end{algorithmic}
\end{algorithm}
\begin{lemma}
{The left and right singular matrices of the effective channel ${\widetilde {\bf{H}}_{{\rm{eq}}}}$ could be directly obtained by applying SVD for each effective sub-channel ${\widetilde {\bf{H}}_{{{ii}}}}, {{i}}=1,2,...,{{P}}$.}
%The search spaces of the selected vectors for different sub-channel don't coincide in the analog precoding design phase.
\end{lemma}
\begin{IEEEproof}
For each effective sub-channel, we have
\begin{equation}\label{3.}
{\widetilde {\bf{H}}_{{ii}}}={\widetilde{{\bf{U }}}}_{{{ii}}}{\widetilde{\bf{\Sigma }}_{{ii}}}{\widetilde{{\bf{V }}}}_{{{ii}}}^{\rm{*}}, {\rm{i}}=1,2,...,{{P}},
\end{equation}
where ${\widetilde{{\bf{U }}}}_{{{ii}}}$ and ${\widetilde{{\bf{V }}}}_{{{ii}}}$ are the left and right singular matrices of ${{\bf{H}}_{{ii}}}$ and ${{\bf{\Sigma }}_{{ii}}}$ is a diagonal matrix with the singular values arranged in decreasing order. Therefore, the effective channel (35) could be written as
\begin{equation}\label{4.}
\begin{split}
{\widetilde {\bf{H}}_{{\rm{eq}}}} & = \left[ \begin{array}{l}
{\widetilde{{\bf{U }}}}_{{\rm{11}}}{\widetilde{\bf{\Sigma }}_{\rm{11}}}{\widetilde{{\bf{V }}}}_{{\rm{11}}}^{\rm{*}}\\
%{\kern 19pt} {\widetilde{{\bf{U }}}}_{{\rm{22}}}{\widetilde{\bf{\Sigma }}_{\rm{22}}}{\widetilde{{\bf{V }}}}_{{\rm{22}}}^{\rm{*}}\\
{\kern 17pt}  \ddots  \\
{\kern 33pt} {\widetilde{{\bf{U }}}}_{{{PP}}}{\widetilde{\bf{\Sigma }}_{{PP}}}{\widetilde{{\bf{V }}}}_{{{PP}}}^{\rm{*}}
\end{array} \right] \\
& ={\widetilde{\bf{U }}}{\widetilde{\bf{\Sigma }}}{\widetilde{{\bf{V }}}}^{\rm{*}},
\end{split}
\end{equation}
where
\begin{equation}\label{4.}
{\widetilde{\bf{U }}} = \left[ \begin{array}{l}
{\widetilde{{\bf{U }}}}_{{\rm{11}}}\\
%{\kern 19pt} {\widetilde{{\bf{U }}}}_{{\rm{22}}}\\
{\kern 17pt}  \ddots  \\
{\kern 33pt} {\widetilde{{\bf{U }}}}_{{{PP}}}
\end{array} \right],
\end{equation}
\begin{equation}\label{4.}
{\widetilde{\bf{\Sigma }}} = \left[ \begin{array}{l}
{{\widetilde{\bf{\Sigma }}}_{{\rm{11}}}}\\
%{\kern 19pt} {{\widetilde{\bf{\Sigma }}}_{{\rm{22}}}}\\
{\kern 17pt}  \ddots  \\
{\kern 33pt} {{\widetilde{\bf{\Sigma }}}_{{{PP}}}}
\end{array} \right],
\end{equation}
\begin{equation}\label{4.}
{\widetilde{{\bf{V }}}}= \left[ \begin{array}{l}
{\widetilde{{\bf{V }}}}_{{\rm{11}}}\\
%{\kern 19pt} {\widetilde{{\bf{V }}}}_{{\rm{22}}}\\
{\kern 17pt}  \ddots  \\
{\kern 33pt} {\widetilde{{\bf{V }}}}_{{{PP}}}
\end{array} \right].
\end{equation}
Since ${\widetilde{{\bf{U }}}}_{{{ii}}}$ and ${\widetilde{{\bf{V }}}}_{{{ii}}}$ are unitary matrices and ${{\widetilde{\bf{\Sigma }}}_{{{ii}}}}$ is a diagonal matrix of non-negative elements, ${\widetilde{\bf{U }}}{\widetilde{\bf{\Sigma }}}{\widetilde{{\bf{V }}}}^{\rm{*}}$ is a singular value decomposition of the channel ${\widetilde {\bf{H}}_{{\rm{eq}}}}$.
\end{IEEEproof}

Note that, since there is no constant magnitude constrains in the digital precoding phase, the digital precoding matrices could be directly obtained by applying SVD. According to Lemma 3, the digital precoding matrices for the transmitter and receiver could be easily determined as
\begin{equation}\label{3.}
{{\bf{F}}_{{\rm{BB}}}} = {\widetilde{{\bf{V }}}}, {{\bf{W}}_{{\rm{BB}}}} = {\widetilde{{\bf{U }}}}.
\end{equation}
Finally, the precoding matrices are normalized to satisfy the power constrains at the transmitter. The proposed scheme is described in detail in {\bf{Algorithm 1}}.

%{\emph{Remark 2:}} In the digital precoding phase, we directly do SVD to each sub-channel which allocate equal power to each paths. Actually, we can utilize water-filling algorithm to allocate the power, which will further improve the performance. Considering the purpose of the paper is to compare the performance of the proposed hybrid precoding scheme with OMP scheme and the full digital precoding scheme, we omit the process of the power allocation.
%%To obtain better performance, we can further utilize more efficient algorithm to get higher spectral efficiency, but obviously the complexity will increase.
%
%{\emph{Remark 3:}} Besides SVD in the digital phase, we can also utilize other methods to obtain the digital matrices, e.g. zero force (ZF) precoding scheme and least square (LS). Moreover, if we assume the digital precoding matrices satisfy ${\bf{F}}_{{\rm{BBi}}}^*{{\bf{F}}_{{\rm{BBi}}}} = {{\bf{I}}_{\rm{Q}}}$, ${\bf{W}}_{{\rm{BBi}}}^*{{\bf{W}}_{{\rm{BBi}}}} = {{\bf{I}}_{\rm{Q}}}$, they can be also obtained by the solution to the corresponding orthogonal Procrustes problem \cite{29Gower}.

%(刚好不需要进行归一化，RF和BB的共轭乘以本身都是单位阵，维度为N_s)
%for each digital precoding cycle, only $Q$ RF chains is used.
\subsection{Computational Complexity Analysis}
In this subsection, we briefly analyze the complexity of proposed HYP-SLD hybrid precoding scheme. To simplify the expression, we assume $Q=Q_1=... = Q_P$ in this subsection. Compared with the OMP scheme, the reduction in complexity is mainly reflected in the following aspects.

{\emph{1) The optimal fully digital precoding is not needed in advance}}. Considering that the optimal precoding matrices converge in chordal distance to antennal response matrices for limited scattering paths \cite{Ayach2012}, we set ${{\bf{A}}_{\rm{ti}}}$ and ${{\bf{A}}_{\rm{ri}}}$ rather than the fully digital precoding matrices as the reference matrices for the $i_{th}$ sub-channel, which could avoid SVD operation of high-dimensional channel matrix.

{\emph{2) The search space of the selected analog precoding vectors is reduced}}. In the analog precoding phase, both the candidate matrix and the reference matrix are divided into $P$ parts according to the spatial lobes. For the $i_{th}$ sub-channel, we only need to select the vector from the corresponding part of the quantization matrix, along which the corresponding part of the reference matrix has the maximum projection.

{\emph{3) The SVD in digital precoding phase is divided}}. After the analog precoding phase, we obtain the digital domain mmWave channel ${{\bf{H}}_{{\rm{eq}}}}$ with $PQ\times PQ$ dimension. Since the effective sub-channels make up a block diagonal matrix ${\widetilde {\bf{H}}_{{\rm{eq}}}}$, we are able to handle each effective $Q\times Q$ sub-channel separately to obtain the digital precoding matrices, which is actually implemented by performing SVD shown as (37).

\begin{table}[!t]
\scriptsize
\renewcommand\arraystretch{2}
\centering
  \caption{THE COMPUTATIONAL COMPLEXITY FOR DIFFERENT HYBRID PRECODING SCHEMES AT THE TRANSMITTER}
\begin{tabular}{|c|c|c|}\hline
\backslashbox{Computation}{Scheme} & OMP  &HYP-SLD\\ \hline
    ${{\bf{F}}_{{\rm{opt}}}}$ &$\mathcal O(N_{\rm t}^2N_{\rm r}+N_{\rm r}^3) $  &NULL \\ \hline
   Analog precoding matrix &$ \mathcal O(2^bN_{\rm t}N_{\rm RF}^{\rm t}N_{\rm s}) $  &$ \mathcal O(2^bN_{\rm t}Q)$\\ \hline
    Digital precoding matrix & $\mathcal O((N_{\rm RF}^{\rm t})^2N_{\rm t}(N_{\rm RF}^{\rm t}+N_{\rm s})) $  &$\mathcal O(PQ^3)$ \\ \hline
  \end{tabular}\label{Table1}
\end{table}

The computation complexities for all hybrid precoding design phases at the transmitter are summarized in Table I. The complexity of computing precoding matrices at both the transmitter and the receiver doubles the number of the operations, while the order of the overall complexity unchanged.
%As we know, the complexity of SVD and the matrix multiplication is $\mathcal O(N^3)$, where $N$ is the matrix size. The division operation in this paper can reduce the overall complexity to $\mathcal O(P\times Q^3)$, which is proportional only to the number of paths in one spatial lobe.
Taking $N_{\rm{t}}=64$, $N_{\rm{r}}=32$, $N_{\rm{RF}}^{\rm{t}}=16$, $N_{\rm{RF}}^{\rm{r}}=8$, $P=4, Q_1=...=Q_P=Q=2, b=7, N_s=PQ=N_{\rm{RF}}^{\rm{r}}$ for example, we find that the reduction of the complexity is more than $99\%$ compared with the OMP scheme.

\section{Proposed Spatial Lobes Diversity Combining Scheme HYP-SLD-MRC}
In the HYP-SLD hybrid precoding scheme, the data streams are associated with the subpaths in each spatial lobe, which inspires us to further utilize these subpaths. As has shown in (39), the singular values matrix of ${\widetilde {\bf{H}}_{{\rm{eq}}}}$ is
\begin{equation}\label{4.}
{\widetilde{\bf{\Sigma }}} = \left[ \begin{array}{l}
{{\widetilde{\bf{\Sigma }}}_{{\rm{11}}}}\\
%{\kern 19pt} {{\widetilde{\bf{\Sigma }}}_{{\rm{22}}}}\\
{\kern 19pt}  \ddots  \\
{\kern 33pt} {{\widetilde{\bf{\Sigma }}}_{{{PP}}}}
\end{array} \right],
\end{equation}
where
\begin{equation}\label{4.}
{{\widetilde{\bf{\Sigma }}}_{{{ii}}}} = \left[ \begin{array}{l}
{{\widetilde{\bf{\Sigma }}}_{{{ii1}}}}\\
%{\kern 19pt} {{\widetilde{\bf{\Sigma }}}_{{{ii2}}}}\\
{\kern 19pt}  \ddots  \\
{\kern 33pt} {{\widetilde{\bf{\Sigma }}}_{{{iiQ_i}}}}
\end{array}\right], i=1,2,...,P,
\end{equation}
contains the singular values for the $i_{th}$ spatial lobe and ${{\widetilde{\bf{\Sigma }}}_{{{ii1}}}}\geq{{\widetilde{\bf{\Sigma }}}_{{{ii2}}}}\geq...\geq{{\widetilde{\bf{\Sigma }}}_{{{iiQ_i}}}}$. It could be observed that ${{\widetilde{\bf{\Sigma }}}_{{{ii}}}}$ has at least one dominated singular value of the total channel matrix $\bf H$ when the AOAs and AODs of different spatial lobes are sufficiently separable. Moreover, when the number of data streams approaches the number of paths, most singular values including the relatively small singular values are used to transmit signals, which causes poor BER performance.

Motivated by this, we determine to design an optional diversity combining scheme to reduce the BER when the IoT system prefers better BER performances. Specifically, we firstly introduce the classic maximal-ratio combining (MRC) diversity combining technique \cite{29MZWin1999}. Then, a new type of diversity combining scheme based on MRC is proposed for the mmWave IoT system.
%\begin{figure}[t]
%\centering
%\includegraphics[scale=0.6]{HYP-SLD-MRC.png}
%\caption{HYP-SLD-MRC}
%\label{HYP-SLD-MRC}
%\end{figure}
\subsection{Maximal-Ratio Combining Scheme}
%We introduce the maximal ratio combining scheme at the receiver in this subsection briefly.
% which conducts a weighted sum across all branches with the objective of maximizing SNR.
In a SIMO system where the receiver is equipped with $N$ antennas, the received signal vector is
\begin{equation}\label{4.25}
{{\bf y_{\rm mrc}} = {\bf h}{s}_{\rm mrc} + \bf{n},}
\end{equation}
where ${\bf{h}} = {[{{{h}}_{\rm{0}}},{{{h}}_{\rm{1}}},...,{{{h}}_{{{N - 1}}}}]^T}$ represents the channel gain vector, ${s}_{\rm mrc}$ is the unit power signal transmitted and $\bf{n}$ is additive white gaussian noise.
MRC conducts a weighted sum across all branches (antennas) with the objective of maximizing SNR \cite{29MZWin1999}, where the weight vector is
%Then the signals from each antennas are added together linearly with a set of weights $\bf{w}$, where
\begin{equation}\label{4.}
{\bf{w}_{\rm mrc}} = {{{{\bf{h}}^ * }}}\big /{\left\|{{\bf{h}}} \right\|}.  %{{{{\left| {\bf{h}} \right|}^2}}}
\end{equation}
The output signal could be obtained by
\begin{equation}\label{4.}
%{\widehat {\bf{y}}}_{\rm mrc} = {\bf{w_{\rm mrc}}{\bf{y_{\rm mrc}}.
{{\bf{\hat y}}_{\rm mrc}} = {{\bf{w}}_{\rm mrc}}{{\bf{y}}_{\rm mrc}}.
\end{equation}
%The purpose of MRC is to maximum the output signal-to-noise ratio (SNR).
Since the signal $\bf{s}_{\rm mrc}$ has unit average power, the instantaneous output SNR could be calculated by
\begin{equation}\label{4.}
\begin{array}{l}
{\rm{\gamma  = }}\dfrac{{{{\left| {{{\bf{h}}^*}{\bf{h}}} \right|}^2}}}{{{\sigma ^2}{{\bf{h}}^*}{\bf{h}}}} = \dfrac{{{{\bf{h}}^*}{\bf{h}}}}{{{\sigma ^2}}} = \sum\limits_{n = 0}^{N - 1} {\dfrac{{{{\left| {{{{h}}_n}} \right|}^2}}}{{{\sigma ^2}}}} = \sum\limits_{n = 0}^{N - 1} {{{\rm{\gamma }}_n}},
\end{array}
\end{equation}
where ${{{\rm{\gamma }}_n}}$ is the input SNR at the $n_{th}$ antenna. As we can see,  the output SNR is the summation of the input SNRs, which is actually the
maximum output SNR.
%which is the sum of the input SNRs at each antenna.
Therefore, the output signal achieves better BER performance due to the increase of the SNR. % and the maximum ratio between signal and noise terms.
Generally, the variable gain weighting factor ${{\bf{w}}_{\rm mrc}}$ could be set to be the ratio of the signal amplitude to the noise power for the diversity path, which has been proved in \cite{29MZWin1999}.
\subsection{Proposed HYP-SLD-MRC Diversity Combining Scheme}
\begin{figure}[t]
\centering
\includegraphics[scale=0.65]{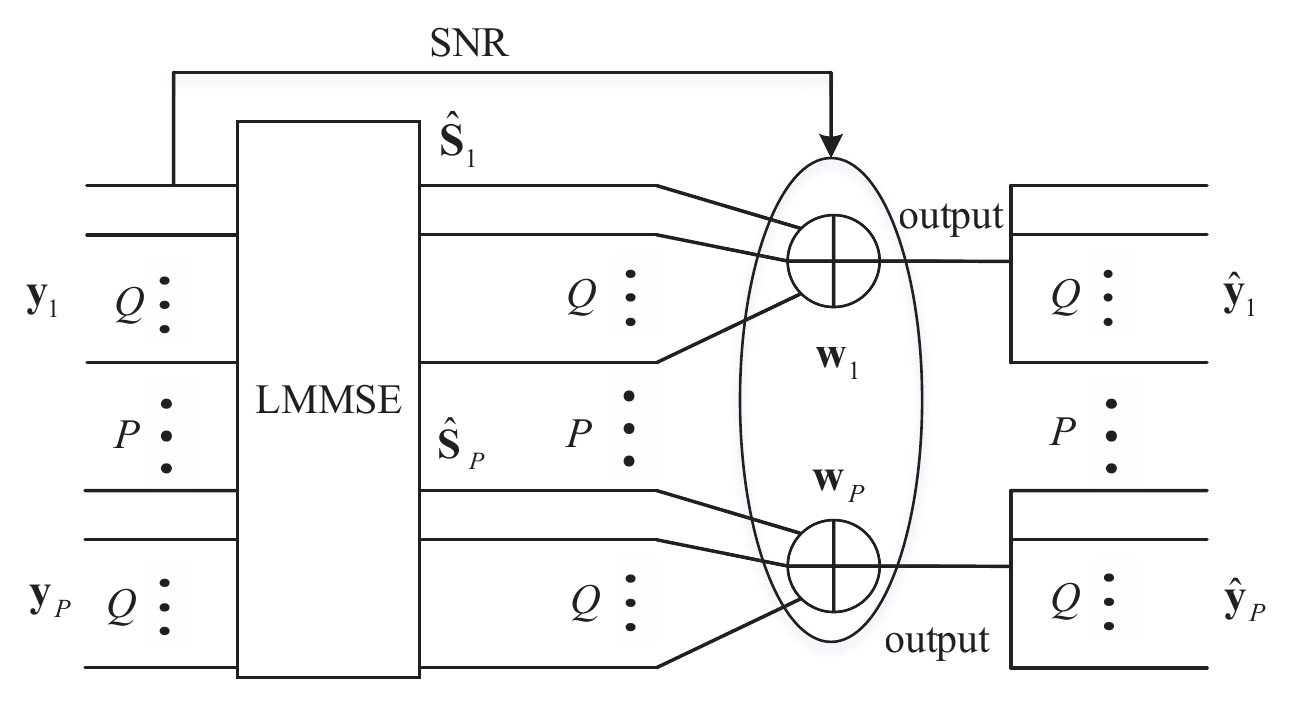}%fig4_HYP-SLD-MRC.eps
\caption{The block diagram of the proposed HYP-SLD-MRC diversity combining scheme.}
\label{HYP-SLD-MRC}
\end{figure}

In this subsection, we demonstrate the proposed HYP-SLD-MRC diversity combining scheme shown in Fig. \ref{HYP-SLD-MRC}. According to the singular values matrix (43), we find that each sub-channel contains at least a dominated singular value of the total channel. Meanwhile, since the number of subpaths satisfies $Q \ll \min ({N_{\rm t}},{N_{\rm r}})$, there is no strong correlation between the data streams on different subpaths. Therefore, we could perform the diversity combining scheme in each sub-channel.

At the transmitter, the signals in (1) are divided into $P$ blocks, which is given by
\begin{equation}\label{4.25}
{\bf{s}} = {[{{\bf{s}}_1},{{\bf{s}}_2},...,{{\bf{s}}_{P}}]^{{T}}},
\end{equation}
where ${{\bf{s}}_{i}}$, ${{i}}=1,2,...,{{P}},$ contains $Q$ copies of one signal transmitted along the $i_{th}$ spatial lobe (sub-channel), i.e.,
\begin{equation}\label{4.26}
{{\bf{s}}_{{i}}} = {[{{{s}}_{{{i1}}}},{{{s}}_{{{i2}}}},...,{{{s}}_{{{iQ_i}}}}]^{{T}}},
\end{equation}
and
\begin{equation}\label{4.27}
{{{s}}_{{{i1}}}}={{{s}}_{{{i2}}}}=...={{{s}}_{{{iQ_i}}}}.
\end{equation}
The above signals are transmitted using the mmWave channel based on the proposed HYP-SLD precoding scheme. Therefore, we associate ${\bf s}_i$ with the $i_{th}$ sub-channel.
%and we make
%\begin{equation}\label{4.27}
%{\widetilde {{s}}_{\rm{i}}} \buildrel \Delta \over = {{{s}}_{{\rm{i1}}}}{\rm{ = }}{{{s}}_{{\rm{i2}}}}{\rm{ = }} \cdots {\rm{ = }}{{{s}}_{{\rm{iQ}}}}.
%\end{equation}
At the receiver, linear minimum mean square error (LMMSE) demodulator is utilized to demodulate the received signal ${\bf{y}}$ in (3). The demodulated signal vector could be obtained by
%${\bf{\hat s}}$
\begin{equation}\label{4.}
%\widehat {\bf{s}} = {({\widehat {\bf{H}}^*}\widehat {\bf{H}} + {\sigma ^2}{\bf{I}})^{ - 1}}{\widehat {\bf{H}}^*}{\bf{y}},
%\begin{array}{l}
\begin{split}
 %&= {({\widehat {\bf{H}}^*}\widehat {\bf{H}} + {\sigma ^2}{\bf{I}})^{ - 1}}{\widehat {\bf{H}}^*}{\bf{y}}\\
 {\bf{\hat s}}&= {({\widehat {\bf{H}}^*}\widehat {\bf{H}} + {\sigma ^2}{\bf{I}})^{ - 1}}{\widehat {\bf{H}}^*}\widehat {\bf{H}}{\bf{s}} + {({\widehat {\bf{H}}^*}\widehat {\bf{H}} + {\sigma ^2}{\bf{I}})^{ - 1}}{\widehat {\bf{H}}^*}{\bf{W}}_{\rm{T}}^{{*}}{\bf{n}}\\
 &={\big[{\hat{\bf{s}}_1},{\hat{\bf{s}}_2},...,{\hat{\bf{s}}_{ P}}\big]^{{T}}}\\
 &={\big[{\hat{{s}}_{11}},{\hat{{s}}_{12}},...,{\hat{{s}}_{ 1Q_1}},...,{\hat{{s}}_{PQ_P}}\big]^{{T}}},
%\end{array}
\end{split}
\end{equation}
where $\widehat {\bf{H}} = {\bf{W}}_{{\rm{BB}}}^*{\bf{W}}_{{\rm{RF}}}^*{\bf{H}}{{\bf{F}}_{{\rm{RF}}}}{{\bf{F}}_{{\rm{BB}}}},$ and ${\hat{\bf{s}}_i}$ contains $Q$ received copies transmitted along the $i_{th}$ sub-channel.

To maximize the output SNR for each sub-channel, we adopt the concept of MRC to combine the $Q$ signal copies of ${\hat{\bf{s}}_{ i}}, {{i}}=1,2,...,{{P}}$. Since the power of the transmitted symbol is normalized, the received signal amplitude after demodulation could be written as
\begin{equation}\label{4.28}
{P_s} = {{{{\left\| {\widehat {\bf{H}}} \right\|}_F^2}}} {\bigg /} {{\left\| {{{\widehat {\bf{H}}}^*}\widehat {\bf{H}} + {\sigma ^2}{\bf{I}}} \right\|}_F}.
\end{equation}
The noise power is
\begin{small}
\begin{equation}\label{4.}
\begin{split}
{P_n} &= {({\widehat {\bf{H}}^*}\widehat {\bf{H}} + {\sigma ^2}{\bf{I}})^{ - 1}}{\widehat {\bf{H}}^*}{\bf{W}}_{\rm{T}}^{\rm{*}}{\bf{n}}*{({({\widehat {\bf{H}}^*}\widehat {\bf{H}} + {\sigma ^2}{\bf{I}})^{ - 1}}{\widehat {\bf{H}}^*}{\bf{W}}_{\rm{T}}^{\rm{*}}{\bf{n}})^*}\\
&= {{{\sigma ^2}{{\left\| {{{\widehat {\bf{H}}}^*}{\bf{W}}_{\rm{T}}^{\rm{*}}} \right\|}_F^2}}}\bigg /{{{{\left\| {{{\widehat {\bf{H}}}^*}\widehat {\bf{H}} + {\sigma ^2}{\bf{I}}} \right\|}_F^2}}}.
\end{split}
\end{equation}
\end{small}
Since the analog precoding matrices are selected from the quantized candidate matrices and the digital precoding matrices are unitary matrices, we have
\begin{equation}\label{4.}
{\bf{W}}_{\rm{T}}^{\rm{*}}{{\bf{W}}_{\rm{T}}}{\rm{ = }}{{\bf{I}}_{{N_s}}}.
\end{equation}
Therefore, (53) could be simplified as
\begin{equation}\label{4.}
{P_n} = {{{\sigma ^2}{{\left\| {{{\widehat {\bf{H}}}^*}} \right\|}_F^2}}}\bigg /{{{{\left\| {{{\widehat {\bf{H}}}^*}\widehat {\bf{H}} + {\sigma ^2}{\bf{I}}} \right\|}_F^2}}}.
\end{equation}
Then, the weight value for the total signals could be computed by
\begin{equation}\label{4.}
\begin{split}
{\widehat{w}} &= \frac{{{P_s}}}{{{P_n}}} = \frac{{\left\| {{{\widehat {\bf{H}}}^*}\widehat {\bf{H}} + {\sigma ^2}{\bf{I}}} \right\|}_F}{{{\sigma ^2}}} \approx \frac{{\left\| {{{\widehat {\bf{H}}}^*}\widehat {\bf{H}}} \right\|}_F}{{{\sigma ^2}}}=\frac{{{{\left\| {\widehat {\bf{H}}} \right\|}_F^2}}}{{{\sigma ^2}}},
\end{split}
\end{equation}
which is actually the SNR before the LMMSE demodulation. Therefore, we set the SNRs of the $Q$ received signal copies before demodulation as the weight values and add the corresponding demodulated signals together, which is shown in Fig \ref{HYP-SLD-MRC}. Each weight value could be calculated as
\begin{equation}\label{4.}
{{{w}}_{ij}} = {{{{\left| {{{{\bf{\tilde s}}}_{ij}}} \right|}^2}}}\big /{{{{\left| {{\bf{\tilde n}}(ij,:)} \right|}^2}}},
\end{equation}
where $1\leq i\leq P, 1\leq j\leq Q_i$, ${\bf{\tilde s}} = \widehat {\bf{H}}{\bf{s}},$ and ${\bf{\tilde n}} = {\bf{W}}_{\rm{T}}^*{\bf{n}}.$
The total weight vector is
\begin{equation}\label{4.}
{\bf{w}} = [{{\bf{{w}}}_{1}},{{\bf{{w}}}_{2}}...,{{\bf{{w}}}_{ P}}],%,...,{\text{{w}}_{{\rm{PQ}}}}],
\end{equation}
where ${{\bf{{w}}}_{i}} = [{{{w}}_{i1}},{{{w}}_{i2}}...,{{{w}}_{i{{Q_i}}}}], i=1,2...,{P},$ represents the weight vector for the $i_{th}$ sub-channel and is normalized in advance. %Each element of ${{\bf{{w}}}_{i}}$ can be calculated as
%\begin{equation}\label{4.}
%{{\rm{w}}_{ij}} = \frac{{{{\left| {{{\hat s}_{ij}}} \right|}^2}}}{{{{\left| {{\bf{n}}(ij,:)} \right|}^2}}}.
%\end{equation}
The output signal vector is
\begin{equation}\label{4.}
\hat {\bf{y}} = {\big[{\hat{\bf{y}}_1},{\hat {\bf{y}}_2},...,\hat {\bf{y}}{}_{{P}}\big]^{{T}}}.
\end{equation}
For each ${\hat {\bf{y}}_i}, i=1,2,...,{\rm P}$, we have
\begin{equation}\label{4.}
{\hat{\bf{y}}_{\rm{i}}} = {\big[{\hat {{y}}_{{{i1}}}},{\hat {{y}}_{{{i2}}}},...,\hat {{y}}{}_{{{iQ_i}}}\big]^{{T}}},
\end{equation}
where each ${\hat {{y}}_{ij}}$ is the linearly combination of the demodulated signals, which could be calculated by
\begin{equation}\label{4.}
\begin{array}{l}
{\hat{ {y}}_{{{i}}1}} = {\hat {{y}}_{{{i2}}}} = ... = {\hat {{y}}_{{{iQ_i}}}} = \sum\limits_{{{j}} = 1}^{Q_i} {{{{w}}_{ij}}{\hat{{s}}_{ij}}}.
\end{array}
\end{equation}
In the proposed HYP-SLD-MRC diversity combining scheme, output SNR is maximized for each sub-channel, which could be easily proved by the Chebyshev inequality. For MRC, the diversity gain is proportional to the number of antennas ($N$) since the output SNR is expanded by $N$ times \cite{32Brennan1959}. Thus, through the proposed HYP-SLD-MRC scheme, $Q$ times diversity gains could be obtained which will improve the BER performance. Moreover, we only need to perform diversity combining on each sub-channel, which makes the signals transmitted along different sub-channels independent. Therefore, $P$ times multiplexing gains could also be obtained.

\section{Simulation Results}
In this section, we evaluate the performances of the proposed HYP-SLD hybrid precoding scheme and HYP-SLD-MRC diversity combining scheme. Both the transmitter and the receiver of IoT devices are equipped with ULA, where $N_{\rm{t}}=64$, $N_{\rm{RF}}^{\rm{t}}=16$ , $N_{\rm{r}}=32$ and $N_{\rm{RF}}^{\rm{r}}=8$ \cite{11Ayach2013}. According to the measurement activity in downtown Manhattan environment \cite{21Samimi2014, 22Samimi2016, 23Rappaport2015}, the frequency of the mmWave is set to be 28 GHz and the bandwidth is set to be 100 MHz. We adopt the clustered narrow-band mmWave channel with sparsity property in the angular domain. According to the step procedures for generating the mmWave channel in \cite{21Samimi2014, 23Rappaport2015}, we make some reasonable simplifications and set the channel parameters as follows.
For $P$ spatial lobes, the whole angular domain is divided into $P$ parts uniformly and the mean angles of spatial lobes ($\widetilde{{\theta}}_i, i=1,2,...,P$) are uniformly distributed within $[0, 2\pi]$, i.e., $\widetilde{{\theta}}_i=\frac{2\pi}{P}(i-1)$. The angle spread of each spatial lobe is set as $\Delta\theta=\frac{\pi}{P}$ to make the angles of paths in different spatial lobes sufficiently separable and the angles of subpaths in one spatial lobe are randomly distributed. The gains of paths in each spatial lobe are assumed to be Rayleigh distributed and the total power of the channel is normalized which satisfies $\mathbb{E}[\left\| {\bf{H}} \right\|_F^2] = {N_{\rm t}}{N_{\rm r}}.$

\begin{figure}
\centering
\subfigure[]{
\centering
\begin{minipage}{0.5\textwidth}
\centering
\includegraphics[scale=0.57]{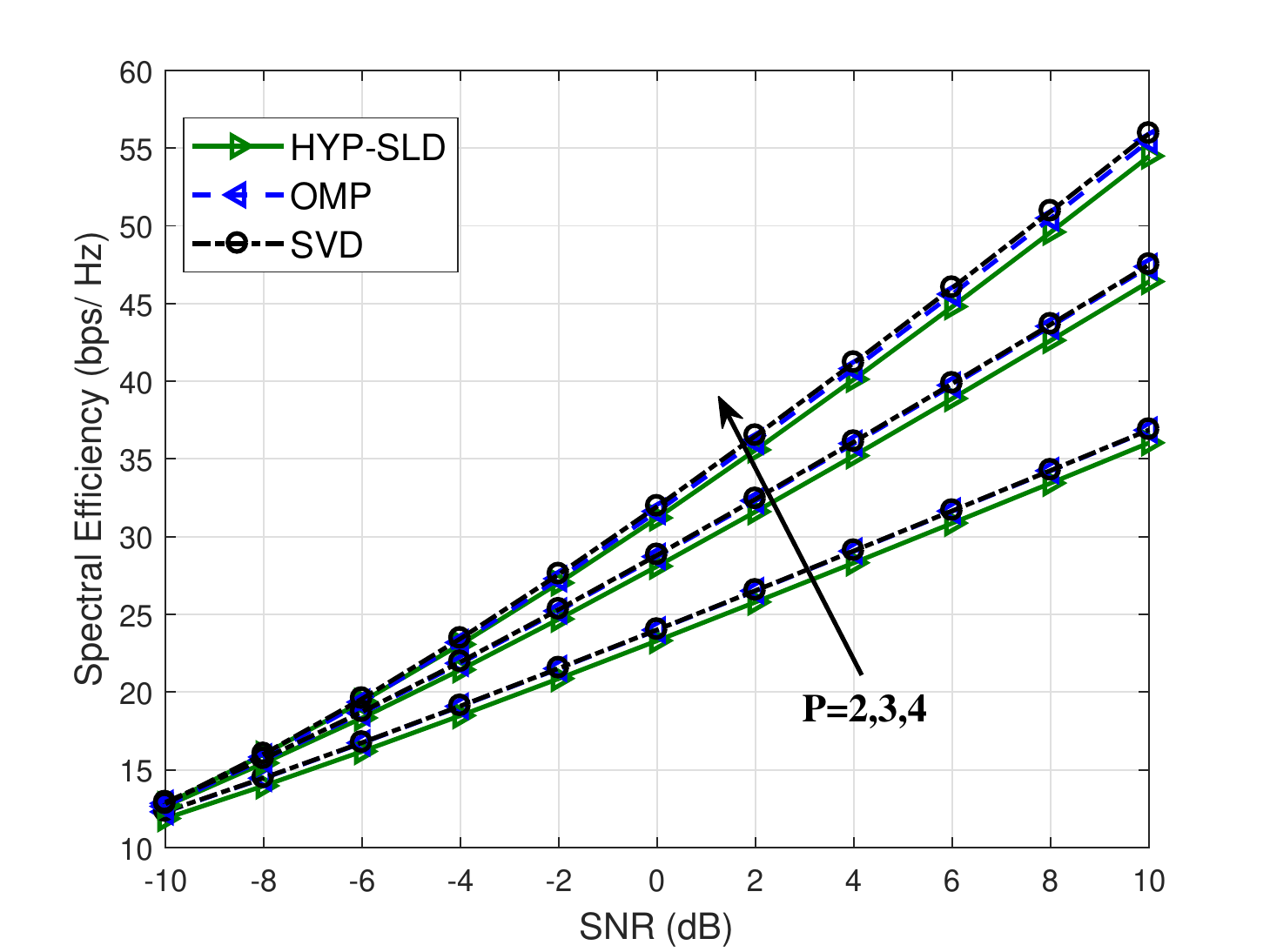}
\end{minipage}
}
\subfigure[]{
\centering
\begin{minipage}{0.5\textwidth}
\centering
\includegraphics[scale=0.57]{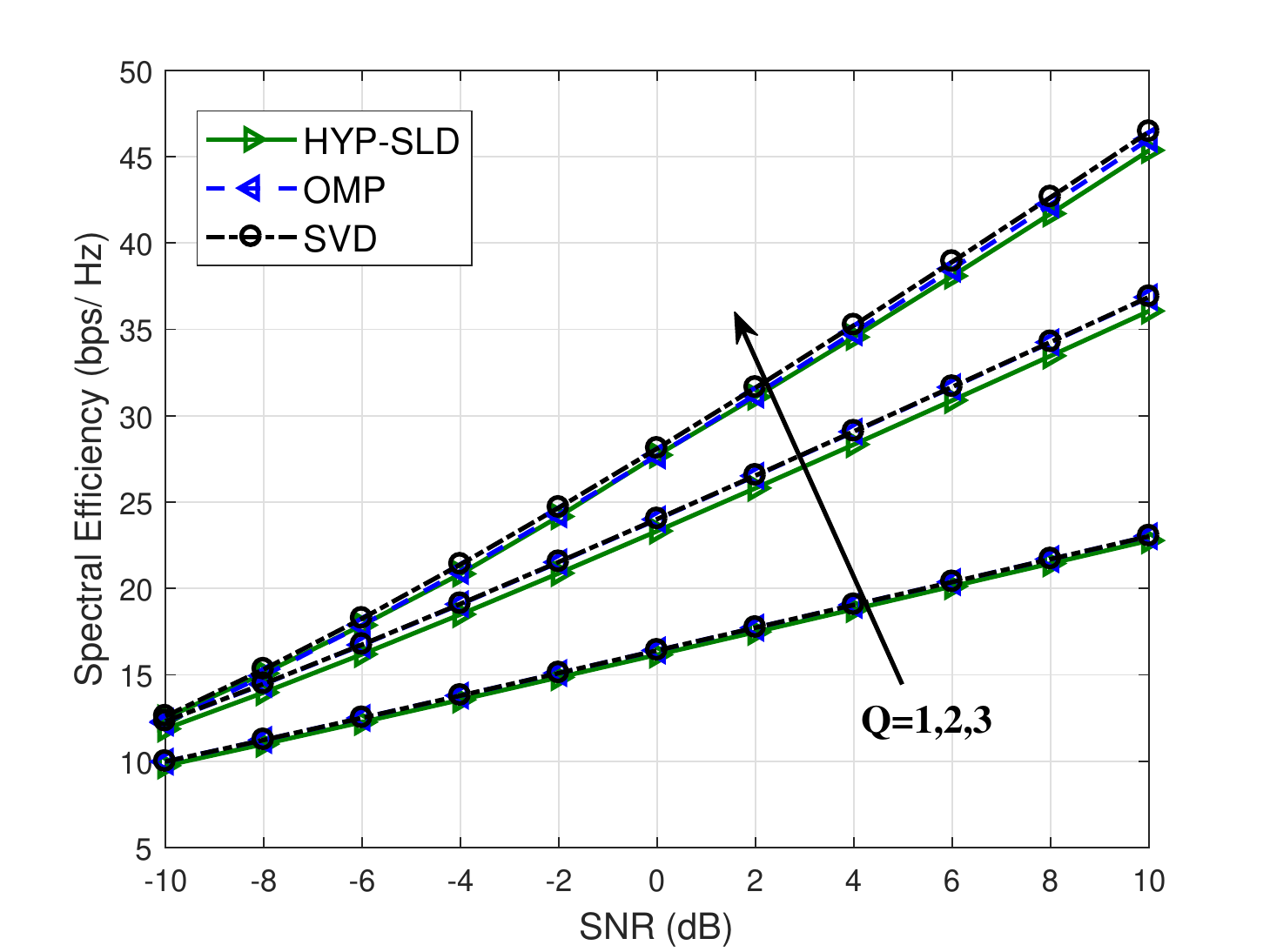}
\end{minipage}
}
\caption{Spectral efficiencies of HYP-SLD, OMP and fully digital precoding schemes with (a) different numbers of spatial lobes $P$ and $Q=2$; (b) different numbers of subpaths $Q$ and $P=2$, where, $N_{\rm{t}}=64,N_{\rm{r}}=32, N_{\rm{RF}}^{\rm{t}}=16, N_{\rm{RF}}^{\rm{r}}=8, b=7.$}
\label{figSprctralEfficiency_sl-subpath}
\end{figure}

Fig. \ref{figSprctralEfficiency_sl-subpath} compares the spectral efficiency of the proposed HYP-SLD, OMP precoding scheme and fully digital precoding scheme (marked as SVD) with different numbers of spatial lobes and subpaths, respectively. Due to the sparse characteristic of mmWave, the number of spatial lobes and subpaths are both very small, specially the maximum number of spatial lobes is 5 for 28 GHz and 73 GHz mmWave signals \cite{22Samimi2016}. It could be observed that the proposed HYP-SLD scheme always achieves similar spectral efficiency as the OMP scheme and the fully digital precoding scheme.

\begin{figure}
\centering
\subfigure[]{
\centering
\begin{minipage}{0.5\textwidth}
\centering
\includegraphics[scale=0.57]{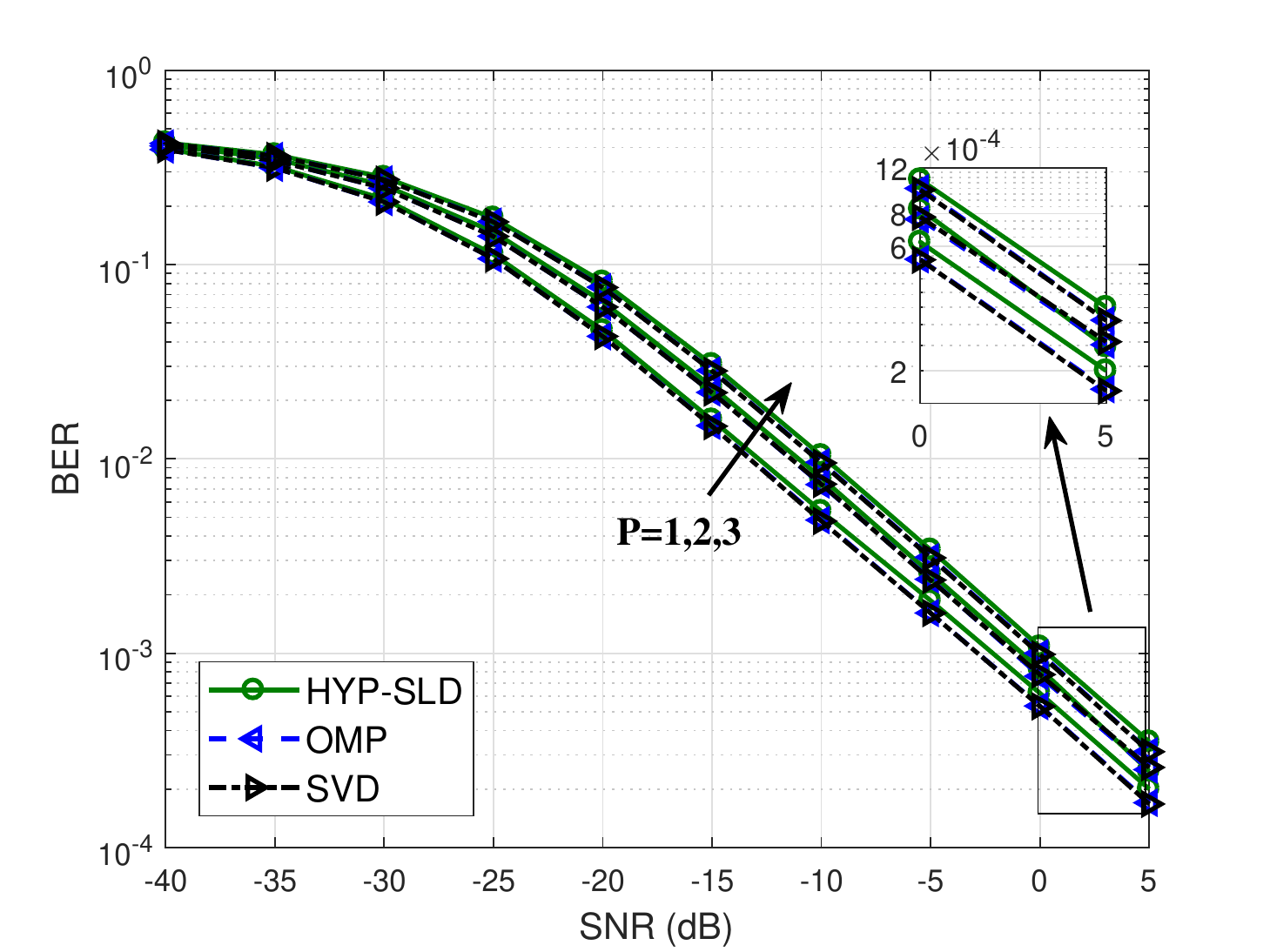}%
\end{minipage}
}
\subfigure[]{
\centering
\begin{minipage}{0.5\textwidth}
\centering
\includegraphics[scale=0.57]{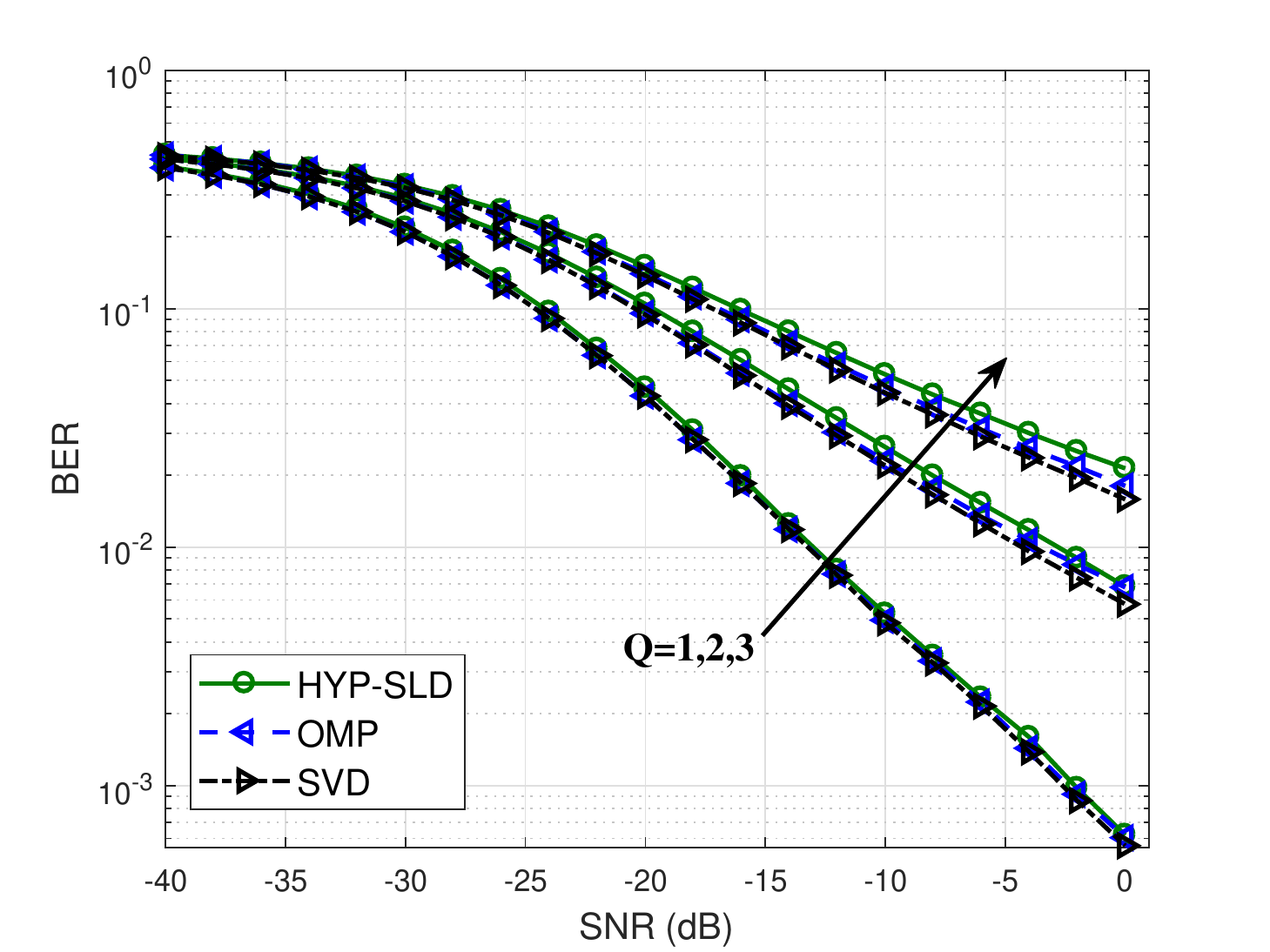}%
\end{minipage}
}

\caption{BERs of HYP-SLD, OMP and fully digital precoding schemes with (a) different numbers of spatial lobes $P$ and $Q=1$; (b) different numbers of subpaths $Q$ and $P=2$,
where $N_{\rm{t}}=64,N_{\rm{r}}=32, N_{\rm{RF}}^{\rm{t}}=16, N_{\rm{RF}}^{\rm{r}}=8, b=7$.}
\label{figBER_sl_subpath}
\end{figure}

In Fig. \ref{figBER_sl_subpath}, we compare the BER performances of HYP-SLD, OMP and the fully digital precoding schemes with different numbers of the subpaths and spatial lobes, respectively. The modulation scheme is QPSK. We observe that three schemes achieve similar BER performances for different $Q$ and $P$. Moreover, the BER performances for different numbers of spatial lobes are very close. However, when the number of the subpaths increases, the BER performances decreases greatly. The above phenomenon demonstrates that the number of subpaths has a greater impact on the BER performance.

\begin{figure}[t]
\centering
\includegraphics[scale=0.57]{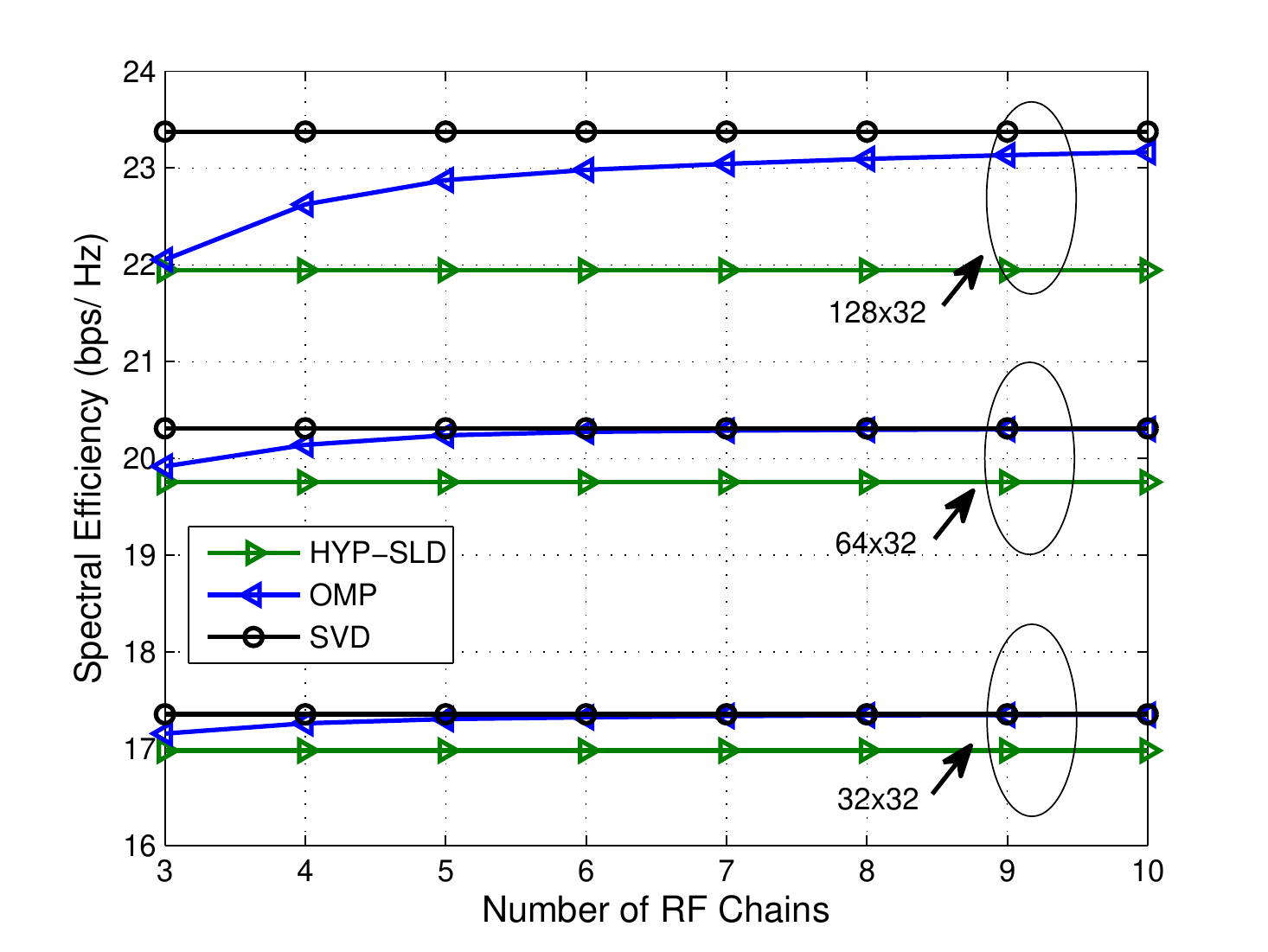}%5000-4-2-RF-r.eps%524_5000-4-2-RF-r.eps
\caption{Spectral efficiencies of HYP-SLD, OMP and fully digital precoding schemes with different numbers of RF chains, where $N_{\rm{s}}=3$, SNR=0 dB, $N_{\rm{RF}}^{\rm{t}}=N_{\rm{RF}}^{\rm{r}}, b=7$.}
\label{fig5000-4-2-RF-r}
\end{figure}
Fig. \ref{fig5000-4-2-RF-r} shows the spectral efficiencies of different schemes with different numbers of RF chains, where SNR = 0 dB, $N_{\rm{s}}=3, b=7$. We observe that when $N_{\rm{RF}}^{\rm{r}}$ varies from 2 to 10, the spectral efficiencies of HYP-SLD remain unchanged. The performance gap originates from two main aspects. 1) We utilize the array response matrices as the reference matrices instead of the optimal precoding matrices; 2) We only utilize $N_{\rm RF}^{\rm r}=N_{\rm RF}^{\rm t}=PQ$ RF chains to transmit and receive signals. Note that, the performance gaps are no more than $5\%$ while the complexity could be reduced by $99\%$.

\begin{figure}[t]
\centering
\includegraphics[scale=0.57]{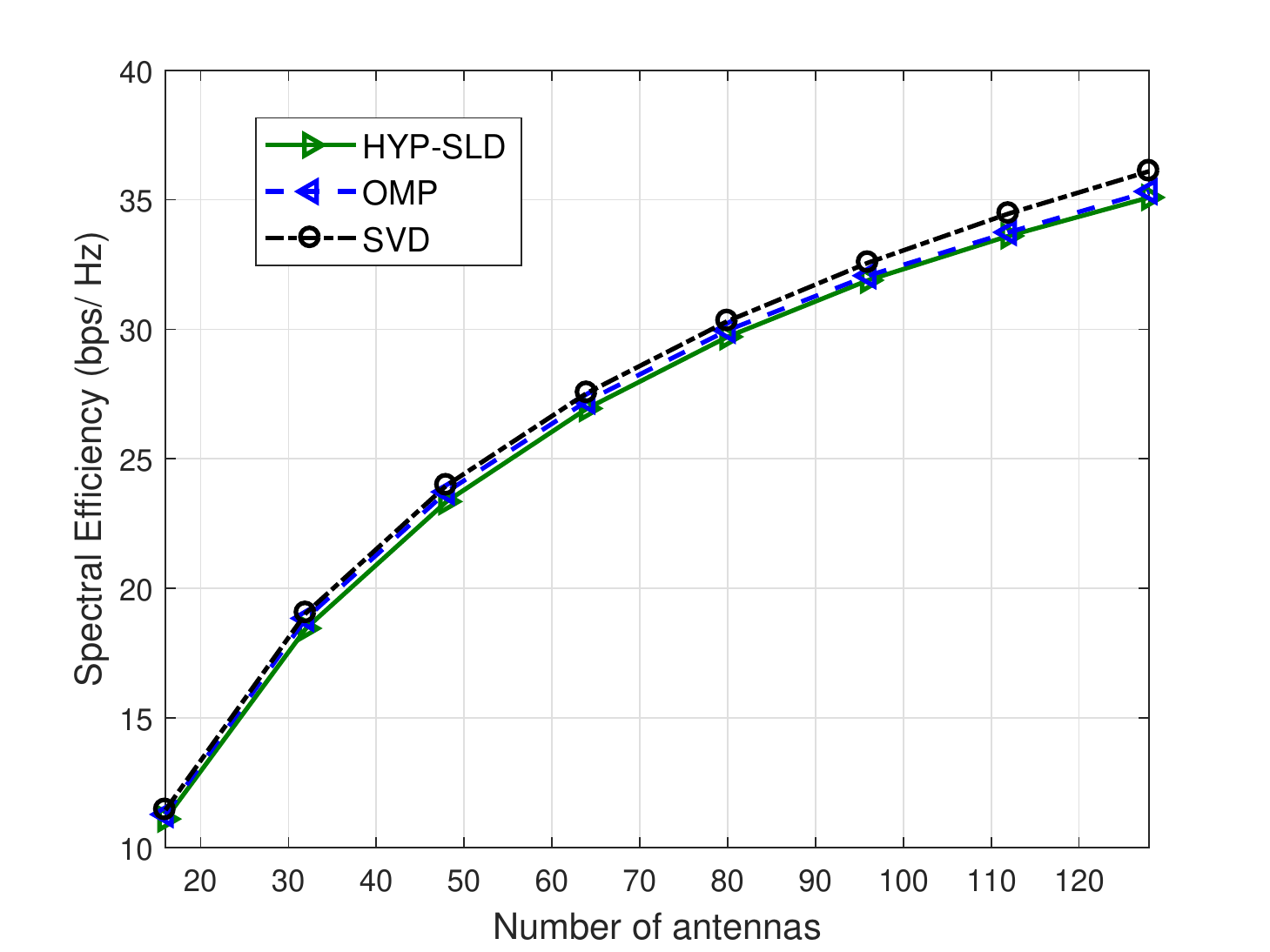}
\caption{Spectral efficiencies of HYP-SLD, OMP and fully digital precoding schemes with different numbers antennas ($N_{\rm t}=N_{\rm{r}}$), where SNR=0 dB, $P=2,Q_i=2,$ $N_{\rm{s}}=N_{\rm{RF}}^{\rm{t}}=N_{\rm{RF}}^{\rm{r}}=4, b=8$.}
\label{fig5000-4-2-RF-antenna}
\end{figure}
Fig. \ref{fig5000-4-2-RF-antenna} compares the spectral efficiencies of different schemes with different numbers of antennas, where $P=Q=2,$ $N_{\rm{s}}=N_{\rm{RF}}^{\rm{t}}=N_{\rm{RF}}^{\rm{r}}=4$, $b=8$ and SNR=0 dB. We observe that the proposed HYP-SLD scheme always achieves similar spectral efficiency as the fully digital precoding scheme even for not very large numbers of antennas (e.g., $N_{\rm t}=N_{\rm r}=16$). In the meantime, it could be observed that when the number of antennas turns very large, there are some performance gaps between the HYP-SLD scheme and the fully digital precoding scheme. This is because the number of quantization bits is not relatively large enough when $N_{\rm t} $ and $N_{\rm{r}}$ become larger.

\begin{figure}[t]
\centering
\includegraphics[scale=0.57]{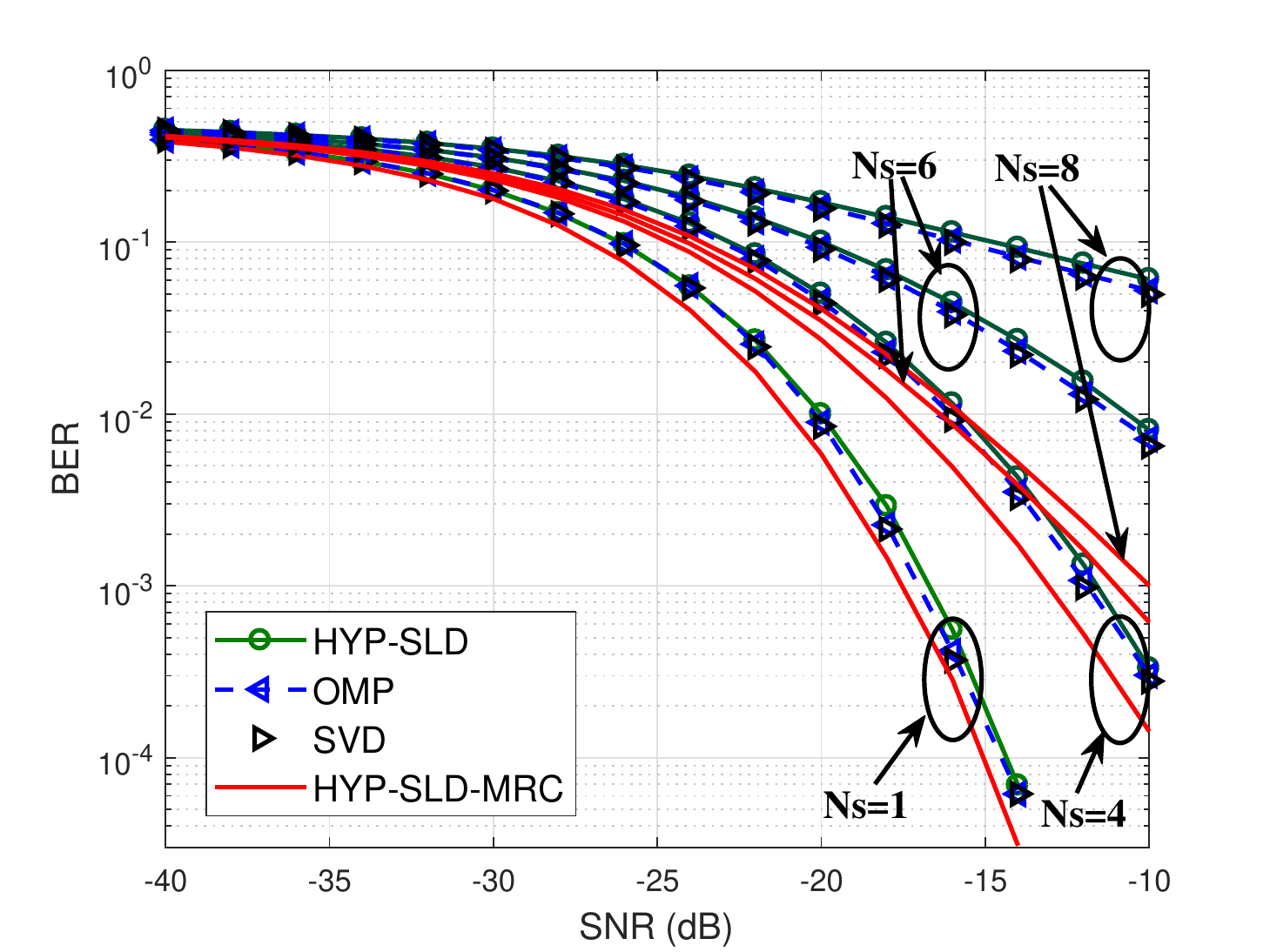}% fig13-BER-datastream-MRC.eps
\caption{BERs of HYP-SLD, OMP, HYP-SLD-MRC and the fully digital precoding schemes with different numbers of data streams, where $N_{\rm{t}}=64,N_{\rm{r}}=32, N_{\rm{RF}}^{\rm{t}}=16, N_{\rm{RF}}^{\rm{r}}=8, P=4, Q_i=2, b=7$.}
\label{figBER_100000_T3_1468_4-2-MRC}
\end{figure}
Fig. \ref{figBER_100000_T3_1468_4-2-MRC} shows the BER performances of the HYP-SLD, OMP, fully digital and the proposed HYP-SLD-MRC diversity combining scheme with different numbers of data streams, where $N_{\rm{t}}=64,N_{\rm{r}}=32, N_{\rm{RF}}^{\rm{t}}=16, N_{\rm{RF}}^{\rm{r}}=8, P=4,Q=2$. It could be observed that the BER performances of the HYP-SLD, OMP, fully digital precoding schemes are almost the same and the BER performances of HYP-SLD-MRC always outperform the fully digital precoding for different numbers of data streams. Moreover, the more the number of data streams is, the more obvious the performance improvement becomes. This is because smaller singular values are used to transmit signals when the number of data streams becomes larger. The proposed HYP-SLD-MRC is able to transmit the signal copy along the smallest singular value and achieves the maximum output SNR for each sub-channel.

\section{Conclusions}
In this paper, we proposed a low complexity hybrid precoding scheme and a diversity combining scheme in the mmWave IoT system. The sparseness property in the angular domain of the mmWave was fully utilized to design the low complexity hybrid precoding scheme. Compared with the widely used OMP scheme, the proposed HYP-SLD greatly reduces the complexity. To improve the BER performance, we proposed a new type of diversity combining scheme to maximize the output SNR for each sub-channel, which allows the diversity gains and the multiplexing gains to be obtained at the same time. Simulation results have demonstrated that the proposed low complexity hybrid precoding scheme exhibits similar spectral efficiency and BER performances as the fully digital precoding scheme. Moreover, the proposed HYP-SLD-MRC achieves significant improvement in BER performance compared with the fully digital precoding scheme. Note that the proposed schemes only concern the single-user narrow-band system. Our future work will focus on multi-user and wide-band scenarios, where the inter-user interference and delay are key points to design the hybrid precoding scheme.

\begin{IEEEbiography}[{\includegraphics[width=1in,height=1.25in]{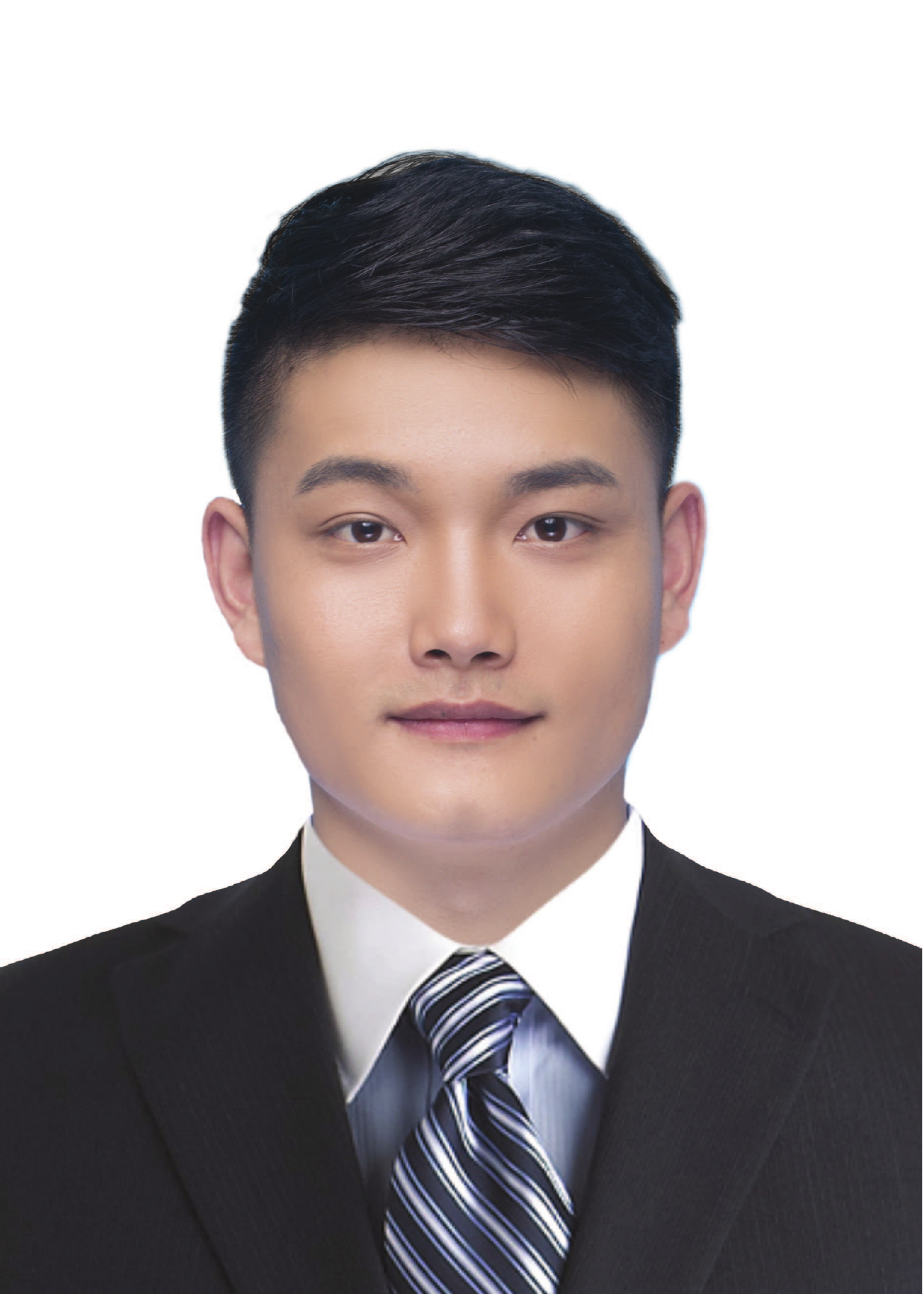}}]
	{Yun Chen } received the B.S. degree from Huazhong University of Science and Technology, Wuhan, P. R. China, in 2016, where he is currently pursuing the Ph.D degree with Wuhan National Laboratory for Optoelectronics and School of Electronic Information and Communications. Since 2018, he has been a Visiting Student with the School of Electronics and Computer Science, University of Southampton, U.K. His current research interests include millimeter wave communications, massive MIMO and FBMC.
\end{IEEEbiography}

\begin{IEEEbiography}[{\includegraphics[width=1in,height=1.25in]{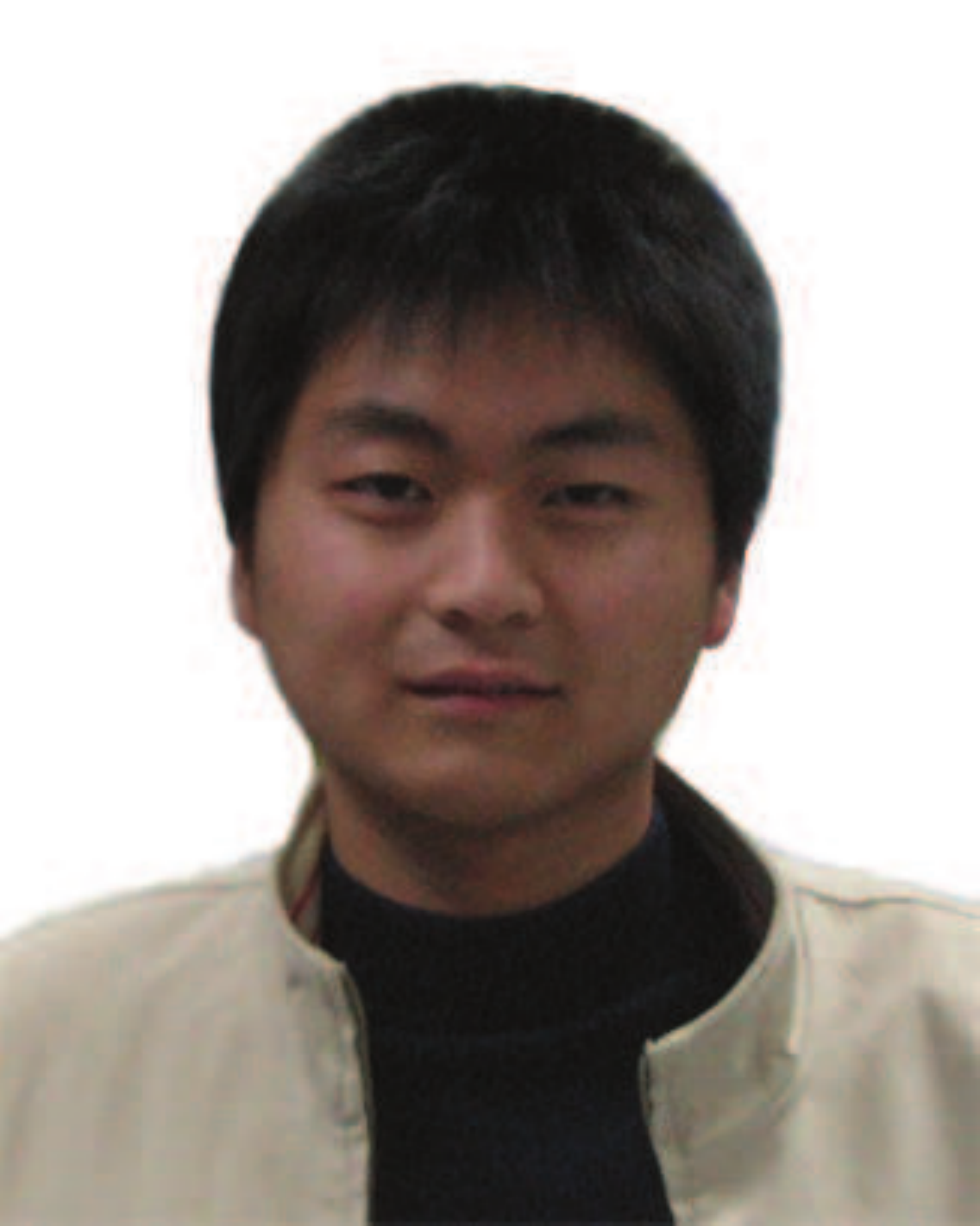}}]
	{Da Chen } received the B.S. and Ph.D. degrees from  Huazhong University of Science and Technology, Wuhan, P. R. China, in 2009 and 2015, respectively. From Sep. 2012 to Aug. 2013, he was a visiting scholar at Northwestern University, USA. From Sep. 2013 to Sep. 2014, he was a visiting scholar at University of Delaware, USA. He is currently an Assistant Professor with the School of Electronics Information and Communications, Huazhong University of Science and Technology, Wuhan, P. R. China. He is serving as an Associate Editor for China Communications. His current research interests include various areas in wireless communications, such as OFDM and FBMC systems.
\end{IEEEbiography}

\begin{IEEEbiography}[{\includegraphics[width=1in,height=1.25in]{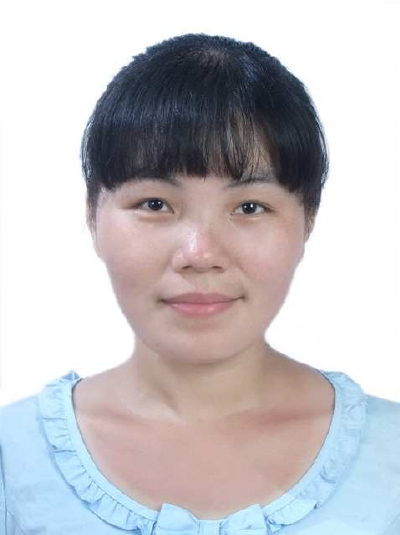}}]
	{Yuan Tian } received the B.S. and M.S. degrees from China University of Geosciences, Wuhan, P. R. China, in 2012 and 2015, respectively. She is currently working towards the Ph.D. degree at Huazhong University of Science and Technology, Wuhan. Her current research interests include various areas in wireless communications, especially for FBMC systems with emphasis on prototype filter design.

\end{IEEEbiography}

\begin{IEEEbiography}[{\includegraphics[width=1in,height=1.25in]{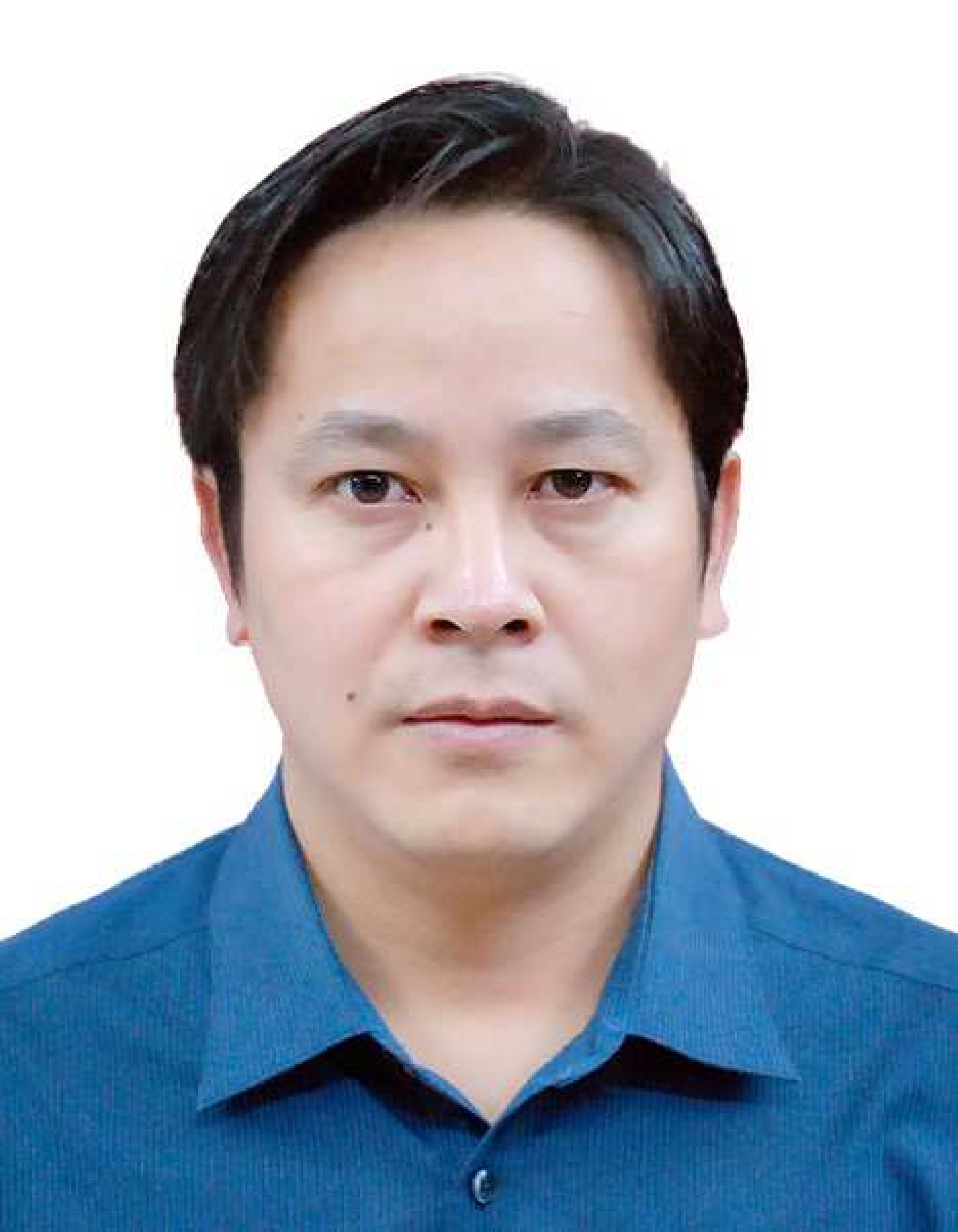}}]
	{Tao Jiang} (M'06-SM'10) received the B.S. and  M.S. degrees in applied geophysics from the China  University of Geosciences, Wuhan, China, in 1997 and 2000, respectively, and the Ph.D. degree in information and communication engineering from Huazhong University of Science and Technology, Wuhan, in 2004. From 2004 to 2007, he was with some universities, such as Brunel University and the University of Michigan at Dearborn. He is currently a Distinguished Professor with the School of Electronics Information and Communications, Huazhong University of Science and Technology. He has authored or co-authored over 300 technical papers in major journals and conferences and nine books/chapters in the areas of communications and networks. He received the NSFC for Distinguished Young Scholars Award in 2013, the Young and Middle-Aged Leading Scientists, Engineers and Innovators by the Ministry of Science and Technology of China in 2014, and the Cheung Kong Scholar Chair Professor by the Ministry of Education of China in 2016. He received the Most Cited Chinese Researchers in Computer Science announced by Elsevier in 2014, 2015, and 2016, respectively. He served or is serving as symposium technical program committee membership of some major IEEE conferences, including INFOCOM, GLOBECOM, and ICC. He is invited to serve as the TPC Symposium Chair for IEEE GLOBECOM 2013, IEEE  WCNC 2013, and ICCC 2013. He is serving as an Associate Editor-in-Chief for China Communications, served or serving as an Associate Editor for some technical journals in communications, including IEEE Transactions on Signal Processing, IEEE Communications Surveys and Tutorials, IEEE Transactions on Vehicular Technology, and IEEE Internet of Things Journal.
\end{IEEEbiography}

\end{document}